\documentclass[journal]{IEEEtran}
\usepackage{tikz}
\usetikzlibrary{decorations.pathmorphing, shapes}
\usetikzlibrary{positioning}
\usetikzlibrary{calc}
\usetikzlibrary{arrows}
 \usepackage{relsize}
 \usepackage[americanvoltage]{circuitikz}

\ifCLASSINFOpdf
\else
\fi
\usepackage{amsmath}
\usepackage{amssymb}
\usepackage{pgfplots}
\usetikzlibrary{calc}
\usepackage[americanvoltage]{circuitikz}
\usepackage{paralist}

\usepackage[caption=false]{subfig}

\newcommand{\vect}[1]{\mathbf{#1}}

\newcommand{\abs}[1]{\left \lvert #1 \right\rvert }
 
\newcommand{\pref}[1]{(\ref{#1})}
\newcommand{\junk}[1] {}

\def\XXint#1#2#3{{\setbox0=\hbox{$#1{#2#3}{\int}$}
\vcenter{\hbox{$#2#3$}}\kern-.5\wd0}}

\newcommand*\widebar[1]{%
  \hbox{%
    \vbox{%
      \hrule height 0.5pt 
      \kern0.3ex
      \hbox{%
        \kern-0.05em
        \ensuremath{#1}%
        \kern-0.05em
      }%
    }%
  }%
}

 \usetikzlibrary{patterns}

\pgfmathsetmacro{\UnitcellWidth}{3.75}
\pgfmathsetmacro{\UnitcellLength}{3.75}
\pgfmathsetmacro{\UnitcellHeight}{3.75}
\pgfmathsetmacro{\ldyl}{3.0}
\pgfmathsetmacro{\ldxl}{3.0}
\pgfmathsetmacro{\tgdcx}{1.6}
\pgfmathsetmacro{\tgdcy}{1.6}
\pgfmathsetmacro{\lcxl}{\ldyl-\tgdcx}
\pgfmathsetmacro{\lcyl}{\ldxl-\tgdcy}
\pgfmathsetmacro{\wdxl}{0.5}
\pgfmathsetmacro{\wdyl}{\wdxl}
\pgfmathsetmacro{\wcxl}{0.3}
\pgfmathsetmacro{\wcyl}{\wcxl}
\pgfmathsetmacro{\h}{0.76}
\pgfmathsetmacro{\hTwo}{-0.76}

\begin{document}

\title{A Macromodeling Approach to Efficiently Compute Scattering from Large Arrays of Complex Scatterers}
\author{Utkarsh~R.~Patel,~\IEEEmembership{Student~Member,~IEEE}, 
        Piero~Triverio,~\IEEEmembership{Senior Member,~IEEE}, and
Sean V.~Hum,~\IEEEmembership{Senior Member,~IEEE}
\\[12pt]
Submitted for publication to the IEEE Transactions on Antennas and Propagation on Dec. 6, 2017.
\thanks{Manuscript received ...; revised ...}%
\thanks{U.~R.~Patel, P.~Triverio, and S. V.~Hum are with the Edward S. Rogers Sr. Department of Electrical and Computer Engineering, University of Toronto, Toronto, M5S 3G4 Canada (email: utkarsh.patel@mail.utoronto.ca, piero.triverio@utoronto.ca, sean.hum@utoronto.ca).}
}

\markboth{Submitted to the IEEE Transactions on Antennas and Propagation}{Submitted to the IEEE Transactions on Antennas and Propagation}%

\maketitle
\begin{abstract}
Full-wave electromagnetic simulations of electrically large arrays of complex antennas and scatterers are challenging, as they consume large amount of memory and require long CPU times.
This paper presents a new reduced-order modeling technique to compute scattering and radiation from large arrays of complex scatterers and antennas.
In the proposed technique, each element of the array is replaced by an equivalent electric current distribution on a fictitious closed surface enclosing the element.
This equivalent electric current density is derived using the equivalence theorem and it is related to the surface currents on the scatterer by the Stratton-Chu formulation.
With the proposed approach, instead of directly solving for the unknown surface current density on the scatterers, we only need to solve for the unknowns on the equivalent surface. 
This approach leads to a reduction in the number of unknowns and better conditioning when it is applied to problems involving complex scatterers with multiscale features.
Furthermore, the proposed approach is accelerated with the adaptive integral equation method to solve large problems.
As illustrated in several practical examples, the proposed method yields speed up of up to 20 times and consumes up to 12 times less memory than the standard method of moments accelerated with the adaptive integral method.
\end{abstract}

\section{Introduction}
Accurate numerical methods are needed to design and optimize large arrays of antennas and scatterers such as phased arrays, frequency selective surfaces, metasurfaces, and reflectarrays.
Currently, most of these arrays are designed and analysed with simulation tools that apply periodic boundary conditions~\cite{Bhattacharyya2006,wan1995,jin2014finite}, neglecting effects of finite array size and dissimilar array elements.
Despite recent advances in computational hardware capabilities, the full-wave electromagnetic simulation of large arrays of antennas and scatterers continues to be a daunting task. 
Unlike volumetric methods such as the finite element method (FEM)~\cite{jin2014finite} and the finite difference (FD) method~\cite{Taflove2005}, the surface integral equation (SIE) method~\cite{Gibson2009} only requires discretization of the surfaces composing the scatterer(s), which makes it an appealing technique to solve many scattering problems.
However, as the electrical size of the problem increases, even the SIE method requires prohibitive amounts of memory and CPU time. 
Multiscale features, commonly found in reflectarrays and metasurfaces, can further hinder the performance of traditional SIE methods.

Electrically large problems can be solved with the SIE method using either acceleration methods or reduced-order modeling methods.
Acceleration and reduced-order modeling techniques achieve scalability in different ways.
In acceleration techniques, the far-field interactions between the basis and testing functions are accelerated using efficient matrix-vector product routines. 
The most commonly used acceleration techniques found in the literature are the fast multipole method (FMM)~\cite{Coifman1993, Greengard1988}, multi-level fast multipole method (MLFMM)~\cite{MLFMM,Gurel2014}, adaptive integral method (AIM)~\cite{Bleszynski96}, pre-corrected fast Fourier transform (pFFT)~\cite{Phillips1997,Zhu2005, Nie2002}, and conjugate gradient fast Fourier transform (CG-FFT)~\cite{Zhuang1996, Wang1998}. 
In FMM and MLFMM, the spherical wave expansion is used to approximate far-field interactions.
Alternatively, the fast Fourier transform may be used to accelerate far-field computations as done in AIM, pFFT, and CG-FFT. 
While the aforementioned acceleration techniques allow simulating large structures, memory consumption and computation time may still increase dramatically in presence of multiscale features.
Multiscale features also require long times to compute near-field interactions which, even in accelerated methods, continue to be a bottleneck.

The goal of reduced-order modeling techniques is to decrease the number of unknowns to be solved in the linear system.
In these techniques, the original set of basis functions is projected onto a new set of basis functions that can well approximate the solution space with a fewer number of basis functions. 
The new set of basis functions is obtained mathematically using either the eigenvalue decomposition or the singular value decomposition.
Common reduced-order modeling techniques in the literature include macro-basis functions~\cite{Sut00}, characteristic basis functions~\cite{Prakash2003}, synthetic basis functions~\cite{Mat07}, and eigencurrent basis functions~\cite{Bekers2009}.

The equivalence principle algorithm~\cite{li2007} is a complementary approach to acceleration and reduced-order modeling techniques for tackling multiscale electromagnetic problems.
In this method, a complex scatterer is enclosed by an equivalent surface and the current distribution on the scatterer is solved iteratively in terms of the tangential electric and magnetic fields on it.
The equivalence principle algorithm can also be hybridized with acceleration algorithms to solve large problems~\cite{Li2008}.

In this paper, we present a novel reduced-order modeling scheme based on the Stratton-Chu formulation and the equivalence theorem to efficiently compute scattering from large arrays of scatterers made up of perfect electric conductors (PECs) in free space.
This work is an extension of a similar idea proposed for a 2D transmission line problem~\cite{TPWRD3}.
In our method, each element of the array is modeled by a so-called macromodel that compactly represents the scattered field from the element.
The macromodel is made up of an equivalent electric current density introduced on a fictitious closed surface surrounding the element and a linear transfer operator to relate the equivalent electric current density to the actual surface currents on the scatterer.
Unlike most SIE methods where both the tangential electric and magnetic fields are expanded with RWG basis functions, we employed RWG and dual RWG basis functions~\cite{Tong2009} to expand the tangential magnetic and electric fields, respectively.
By using RWG and dual RWG basis functions all integral operators in the Stratton-Chu formulation can be well-tested, which ultimately allows us to derive a macromodel that is robust.
The proposed method is faster than traditional SIE methods for three reasons.
First, the proposed method has fewer unknowns than the original problem.
In the original problem, the unknowns are coefficients of electric surface current density on the scatterer. 
On the other hand, in the equivalent problem, the unknowns are only on the equivalent surface. 
Hence, for complex scatterers with multiscale features, the number of unknowns on the equivalent surface could be significantly lower than the number of unknowns on the scatterer.
Second, the macromodel approach improves the condition number of the linear system since the unknowns are only the equivalent surface, which has no fine features.
Finally, the proposed approach exploits repeatability of elements in the array. 
That is, the macromodels generated for one element can be reused for other identical elements in the array, which leads to significant computational and memory savings.
In comparison to the equivalence principle algorithm~\cite{li2007}, the proposed method only requires a single equivalent current source to model each element~\cite{AWPL2017}, which results in an overall simpler formulation with fewer integral operators and unknowns.

The paper is organized as follows. 
First, we provide the mathematical framework to create the macromodel for an element of the array in Sec.~\ref{sec:macromodel_single}.
Then, in Sec.~\ref{sec:macromodel_multiple}, we replace all elements of the array by their macromodels.
Once macromodels are generated, the coupling between them is captured by the electric field integral equation in Sec.~\ref{sec:exterior}.
To tackle electrically large problems, the proposed approach is accelerated with AIM in Sec.~\ref{sec:aim}.
In Sec.~\ref{sec:Results}, we  present three numerical examples to show the accuracy and efficiency of the proposed method.
Finally, Sec.~\ref{sec:conclusion} presents concluding remarks on this work.

\begin{figure}[t]
\null \hfill
\subfloat[\label{fig:sample_cell_equivalent_1} Original] {\input{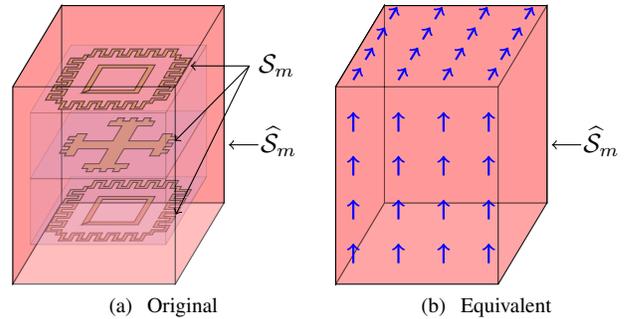}}
\hfill 
\subfloat[\label{fig:sample_cell_equivalent_2} Equivalent] {\begin{tikzpicture}[xscale= 0.3, yscale =0.5]
\pgfmathsetmacro{\tgdcx}{1.6}
\pgfmathsetmacro{\tgdcy}{1.6}
\pgfmathsetmacro{\lcxl}{\ldyl-\tgdcx}
\pgfmathsetmacro{\lcyl}{\ldxl-\tgdcy}
\pgfmathsetmacro{\wdxl}{0.5}
\pgfmathsetmacro{\wdyl}{\wdxl}
\pgfmathsetmacro{\wcxl}{0.3}
\pgfmathsetmacro{\wcyl}{\wcxl}
\pgfmathsetmacro{\h}{0.76}
\pgfmathsetmacro{\hTwo}{-0.76}
\pgfmathsetmacro{\InnerLoopWin}{2.0}
\pgfmathsetmacro{\InnerLoopWout}{2.5}
\pgfmathsetmacro{\mh}{0.7}
\pgfmathsetmacro{\ml}{0.45}
\pgfmathsetmacro{\mw}{0.17}
\pgfmathsetmacro{\numgroove}{5}
\pgfmathsetmacro{\UnitcellWidth}{6.0}
\pgfmathsetmacro{\UnitcellLength}{6.0}
\pgfmathsetmacro{\UnitcellHeight}{8.0}
\pgfmathsetmacro{\epsthres}{0.0001}
\pgfmathsetmacro{\h}{1.75}
\pgfmathsetmacro{\xslantval}{1}
\pgfmathsetmacro{\xscaleval}{1}
\pgfmathsetmacro{\yscaleval}{0.3}
\pgfmathsetmacro{\numarrow}{3}
\pgfmathsetmacro{\numarrowa}{4}



\begin{scope}[shift = {($(0,-1.5*\h)$)}, every node/.append style={xslant=\xslantval},xslant=\xslantval, xscale= \xscaleval, yscale = \yscaleval]
\coordinate (p1) at (-\UnitcellWidth*0.6,-\UnitcellLength*0.6);
\coordinate (p2) at (\UnitcellWidth*0.6,-\UnitcellLength*0.6);
\coordinate (p3) at (\UnitcellWidth*0.60, \UnitcellLength*0.6);
\coordinate (p4) at (-\UnitcellWidth*0.60,\UnitcellLength*0.6);
\draw[fill=red!40!white,opacity=0.4] (p1)--(p2)--(p3)--(p4)--(p1);
\end{scope}
\draw[fill=red!40!white, opacity = 1] (p4) -- (p3) -- ($(p3) + (0,3*\h)$) -- ($(p4) + (0,3*\h)$) -- (p4);
\draw[fill=red!40!white, opacity = 1] (p4) -- (p1) -- ($(p1) + (0,3*\h)$) -- ($(p4) + (0,3*\h)$) -- (p4);

\draw[fill=red!40!white, opacity = 0.4] (p1) --++ (0,3*\h) --++ (1.2*\UnitcellWidth,0) -- (p2) -- (p1);
\draw[fill=red!40!white, opacity = 0.4] (p2) -- (p3) -- ($(p3) + (0,3*\h)$) -- ($(p2) + (0,3*\h)$) -- (p2);

\begin{scope}[shift = {($(0,1.5*\h)$)}, every node/.append style={xslant=\xslantval},xslant=\xslantval, xscale= \xscaleval, yscale = \yscaleval]
\coordinate (p1) at (-\UnitcellWidth*0.6,-\UnitcellLength*0.6);
\coordinate (p2) at (\UnitcellWidth*0.6,-\UnitcellLength*0.6);
\coordinate (p3) at (\UnitcellWidth*0.6, \UnitcellLength*0.6);
\coordinate (p4) at (-\UnitcellWidth*0.6,\UnitcellLength*0.6);
\draw[fill=red!40!white,opacity=0.4] (p1)--(p2)--(p3)--(p4)--(p1);
\end{scope}

\begin{scope}[shift = {($(0,-1.5*\h)$)}, every node/.append style={xslant=\xslantval},xslant=\xslantval, xscale= \xscaleval, yscale = \yscaleval]
\coordinate (p1) at (-\UnitcellWidth*0.6,-\UnitcellLength*0.6);
\coordinate (p2) at (\UnitcellWidth*0.6,-\UnitcellLength*0.6);
\coordinate (p3) at (\UnitcellWidth*0.60, \UnitcellLength*0.6);
\coordinate (p4) at (-\UnitcellWidth*0.60,\UnitcellLength*0.6);
\draw[fill=red!40!white,opacity=0.4] (p1)--(p2)--(p3)--(p4)--(p1);
\end{scope}
\draw[fill=red!40!white, opacity = 1] (p4) -- (p3) -- ($(p3) + (0,3*\h)$) -- ($(p4) + (0,3*\h)$) -- (p4);
\draw[fill=red!40!white, opacity = 1] (p4) -- (p1) -- ($(p1) + (0,3*\h)$) -- ($(p4) + (0,3*\h)$) -- (p4);

\draw[fill=red!40!white, opacity = 0.4] (p1) --++ (0,3*\h) --++ (1.2*\UnitcellWidth,0) -- (p2) -- (p1);
\foreach \x in {0,...,\numarrow}
{
\foreach \y in {0,...,\numarrow}
{
	\coordinate (t1) at (-\UnitcellWidth*0.65 + \x*1.0*\UnitcellWidth/\numarrow, -1.8*\h +\y*1.0*2*\h/\numarrow);
	\coordinate (t2) at (-\UnitcellWidth*0.65 + \x*1.0*\UnitcellWidth/\numarrow, -1.8*\h + \y*1.0*2*\h/\numarrow + 0.30*\h);
\draw [blue, ->,thick] (t1) -- (t2);
}
}

\draw[fill=red!40!white, opacity = 0.4] (p2) -- (p3) -- ($(p3) + (0,3*\h)$) -- ($(p2) + (0,3*\h)$) -- (p2);

\begin{scope}[shift = {($(0,1.5*\h)$)}, every node/.append style={xslant=\xslantval},xslant=\xslantval, xscale= \xscaleval, yscale = \yscaleval]
\coordinate (p1) at (-\UnitcellWidth*0.6,-\UnitcellLength*0.6);
\coordinate (p2) at (\UnitcellWidth*0.6,-\UnitcellLength*0.6);
\coordinate (p3) at (\UnitcellWidth*0.6, \UnitcellLength*0.6);
\coordinate (p4) at (-\UnitcellWidth*0.6,\UnitcellLength*0.6);
\draw[fill=red!40!white,opacity=0.4] (p1)--(p2)--(p3)--(p4)--(p1);
\foreach \x in {0,...,\numarrow}
{
\foreach \y in {0,...,\numarrow}
{
	\coordinate (t1) at (-\UnitcellWidth*0.5 + \x*1.0*\UnitcellWidth/\numarrow, -\UnitcellLength*0.50 + \y*0.9*\UnitcellLength/\numarrow);
	\coordinate (t2) at (-\UnitcellWidth*0.5 + \x*1.0*\UnitcellWidth/\numarrow, -\UnitcellLength*0.50 + \y*0.9*\UnitcellLength/\numarrow + 0.18*\UnitcellLength);
\draw [blue, ->, thick] (t1) -- (t2);
}
}

\end{scope}
\node at (7.1,0.1) {$\widehat{\cal S}_m$};
\draw[->] (6.2, 0) -- (4.9, 0);
\end{tikzpicture}}
\hfill \null
\caption{(a): Original configuration: Sample unit cell made up of metallic scatterers enclosed by the equivalent surface. (b): Equivalent configuration where the metallic scatterers are removed from the equivalent surface. To restore the fields outside $\widehat{\cal S}_m$, an equivalent electric current density (shown with blue arrows) is introduced on $\widehat{\cal S}_m$.}
\label{fig:equivalent_cell}
\end{figure}

\section{Macromodel of a Single Element}
\label{sec:macromodel_single}

We consider the problem of computing scattering from an $M$-element array of complex scatterers.
In this section, we derive a macromodel for one element of the array. 
The macromodel derived in this section is based on the equivalence theorem and it is therefore exact, except for numerical errors introduced by discretization of fields and currents. 
The derived macromodel efficiently describes the electromagnetic behaviour of the original element using fewer unknowns, reducing memory consumption and computation time.
For simplicity, we assume that the structure is excited by an electric field incident on the scatterer. 
Results in Sec.~\ref{sec:Results}, however, show that the proposed idea is also applicable to driven antenna elements. 

\subsection{Fields and Currents Discretization}

We consider the $m$-th element of the array.
This element consists of several PEC surfaces, which are denoted by ${\cal S}_m$. 
We enclose the element by a fictitious closed surface, which is denoted by $\widehat{\cal S}_m$. 
A sample scatterer and the enclosing equivalent surface are shown in Fig.~\ref{fig:sample_cell_equivalent_1}.

\subsubsection{Discretization of ${\cal S}_m$}

The electric surface current density on the metallic scatterers is expanded as
\begin{equation}
\vect{J}_{m}(\vect{r}) = \sum_{n=1}^{N_m} J_{m,n} {\bf \Lambda}_{m,n}(\vect{r}) \,,
\label{eq:J_expand}
\end{equation}
where ${\bf \Lambda}_{m,n}(\vect{r})$ is the $n$-th RWG basis function~\cite{RWG} on the $m$-th element. 
The coefficients of $\vect{J}_m(\vect{r})$ in~\eqref{eq:J_expand} are collected into vector
\begin{equation}
\mathbb{J}_m = \begin{bmatrix} J_{m,1} & J_{m,2} & \hdots & J_{m, N_m}\end{bmatrix}^T \,.
\label{eq:J_expand2}
\end{equation}

\subsubsection{Discretization of $\widehat{\cal S}_m$}
Like the surface current density on the element, the tangential magnetic field on the equivalent surface $\widehat{\cal S}_m$ is also expanded with RWG basis functions
\begin{equation}
\vect{n} \times \widehat{\vect{H}}_m(\vect{r}) = \sum_{n=1}^{\widehat{N}_m} \widehat{H}_{m,n} \widehat{\bf \Lambda}_{m,n}(\vect{r})\,,
\label{eq:Hhat_expand}
\end{equation}
where $\vect{n}$ is the normal vector pointing into the surface $\widehat{\cal S}_m$.
Note that we use $\,\widehat{ }\,$ to denote all quantities on the equivalent surface.
The tangential electric field on $\widehat{\cal S}_m$ is, instead, expanded with dual RWG basis functions~\cite{Tong2009, Chen1990, Andriulli2008} 
\begin{equation}
\vect{n} \times \widehat{\vect{E}}_m(\vect{r}) = \sum_{n=1}^{\widehat{N}_m} \widehat{E}_{m,n} \widehat{\bf \Lambda}_{m,n}'(\vect{r})\,.
\label{eq:Ehat_expand}
\end{equation}
The $n$-th dual RWG basis function $\widehat{\bf \Lambda}_{m,n}'$ is approximately orthogonal to the $n$-th RWG basis function $\widehat{\bf \Lambda}_{m,n}$. 
The use of both RWG functions and their duals is necessary to achieve a well-conditioned formulation and high robustness, as will be discussed in detail in the next sections.
As in \eqref{eq:J_expand2}, we collect the coefficients of the tangential electric and magnetic fields in~\eqref{eq:Hhat_expand} and~\eqref{eq:Ehat_expand} into vectors
\begin{align}
\mathbb{\widehat{H}}_m &= \begin{bmatrix} \widehat{H}_{m,1} & \widehat{H}_{m,2} & \hdots &\widehat{H}_{m, \widehat{N}_m}\end{bmatrix}^T \\
\mathbb{\widehat{E}}_m &= \begin{bmatrix} \widehat{E}_{m,1} & \widehat{E}_{m,2} & \hdots &\widehat{E}_{m, \widehat{N}_m}\end{bmatrix}^T\,.
\end{align}

\subsection{Equivalence Theorem}
\label{sec:equivalence_overview}
As seen from Fig.~\ref{fig:sample_cell_equivalent_1}, a unit cell of a typical metasurface or reflectarray can be quite complex.
In order to handle complex unit cells efficiently, we apply the equivalence theorem~\cite{Bal05} to $\widehat{\cal S}_m$.
As shown in Fig.~\ref{fig:sample_cell_equivalent_2}, we replace all PECs inside the surface with free space and introduce on $\widehat{\cal S}_m$ an equivalent electric current density~\cite{Bal05}
\begin{equation}
\widehat{\vect{J}}_m(\vect{r}) = \vect{n} \times \left[ \widetilde{\mathbf{H}}_m(\vect{r}) -  \widehat{{\mathbf{H}}}_m(\vect{r}) \right]\,
\label{eq:JEquivalence2}
\end{equation}
and an equivalent magnetic current density~\cite{Bal05}
\begin{equation}
\widehat{\vect{M}}_m(\vect{r}) = -\vect{n} \times \left[ \widetilde{\mathbf{E}}_m(\vect{r}) -  \widehat{{\mathbf{E}}}_m(\vect{r}) \right]\,.
\label{eq:MEquivalence2}
\end{equation}
In~\eqref{eq:JEquivalence2} and~\eqref{eq:MEquivalence2}, $\widetilde{\mathbf{H}}_m(\vect{r})$ and $\widetilde{\mathbf{E}}_m(\vect{r})$ are the electric and magnetic fields on $\widehat{\cal S}_m$ in the equivalent problem. 
According to the equivalence theorem, these currents will produce the same electric and magnetic fields outside $\widehat{\cal S}_m$ as the actual currents on the PEC elements, allowing us to compute the radiation from the array.

Most SIE formulations are based on the Love's equivalence theorem~\cite{Bal05}, which sets $\widetilde{\vect{H}}_m$ and $\widetilde{\vect{E}}_m$ to zero, and require both $\widehat{\vect{J}}_m(\vect{r})$ and $\widehat{\vect{M}}_m(\vect{r})$ to restore the electromagnetic fields outside the scatterer.
However, in this work, we \emph{enforce} that $\widetilde{\vect{E}}_m(\vect{r})$ is equal to  $\widehat{\vect{E}}_m(\vect{r})$~\cite{DeZ05,TMTT1}. Therefore, the magnetic equivalent current in~\eqref{eq:MEquivalence2} is zero and only a single equivalent current source is required to model the scatterer.
This single-source equivalence approach has been successfully applied to model 3D dielectrics~\cite{AWPL2017, APS2017} and conductors~\cite{EPEPS2017}.

We expand the tangential magnetic field $\vect{n} \times \widetilde{\vect{H}}_m(\vect{r})$ on $\widehat{\cal S}_m$ in the equivalent problem using RWG basis functions
\begin{equation}
\vect{n} \times \widetilde{\vect{H}}_m(\vect{r}) = \sum_{n=1}^{\widehat{N}_m} \widetilde{H}_{m,n} \widehat{\mathbf{\Lambda}}_{m,n}(\vect{r}) \,,
\label{eq:Htildehat_expand1}
\end{equation}
and collect its expansion coefficients into vector
\begin{equation}
\mathbb{\widetilde{H}}_m = \begin{bmatrix} \widetilde{H}_{m,1} & \widetilde{H}_{m,2} & \hdots &\widetilde{H}_{m, \widehat{N}_m}\end{bmatrix}^T\,. \\
\end{equation}
Similarly, the equivalent electric current density is also expanded with RWG basis functions as
\begin{equation}
\widehat{\mathbf{J}}_{m} (\vect{r}) = \sum_{n=1}^{\widehat{N}_m} \widehat{J}_{m,n} \widehat{\bf \Lambda}_{m,n}(\vect{r}) \,,
\label{eq:Jeq1}
\end{equation}
and its expansion coefficients are collected into vector
\begin{equation}
\widehat{\mathbb{J}}_m = \begin{bmatrix} \widehat{J}_{m,1} & \widehat{J}_{m,2} & \hdots & \widehat{J}_{m,\widehat{N}_m} \end{bmatrix}^T \,.
\label{eq:Jeq2_expansion}
\end{equation}
By substituting,~\eqref{eq:Jeq1},~\eqref{eq:Htildehat_expand1} and~\eqref{eq:Hhat_expand} into~\eqref{eq:JEquivalence2} we obtain
\begin{equation}
\widehat{\mathbb{J}}_{m} = \widetilde{\mathbb{H}}_m - \widehat{\mathbb{H}}_m \,
\label{eq:Jeq2}
\end{equation}
in the discrete domain.

Next, we simplify~\eqref{eq:Jeq2} by applying the Stratton-Chu formulation to two problems: the original problem and the equivalent problem.

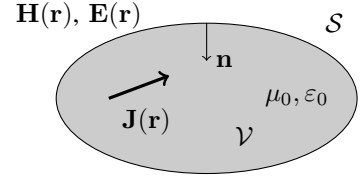
\begin{figure}[t]
\centering
\usetikzlibrary{arrows, decorations.markings}

\begin{tikzpicture}[scale=1, every node/.style={scale=1}]

\draw[black, fill=black!20] (0,0) ellipse (2cm and 1cm);
\node at (0.5,-0.5) {${\cal V}$};
\node at (1.7, 1) {${\cal S}$};
\draw [->] (0,1) -- (0,0.5);
\node at (0, 0.5) [right] {{$\vect{n}$}};
\node at (-1.7, 1.1) {$\vect{H}(\vect{r})$, $\vect{E}(\vect{r})$};
\draw[->, very thick]  (-1.3, 0) -- (-0.5,0.3);
\node at (-0.8, 0.0) [below] {$\vect{J}(\vect{r})$};
\node at (1.2,0) {$\mu_0, \varepsilon_0$};

\end{tikzpicture}
\caption{Sample boundary value problem considered in Sec.~\ref{sec:Stratton}. Electric and magnetic fields are defined on the boundary of a closed surface ${\cal S}$. There is also an additional electric current density $\vect{J}(\vect{r})$ inside ${\cal V}$. }
\label{fig:stratton}
\end{figure}

\subsection{Stratton-Chu Formulation}
\label{sec:Stratton}
On a generic closed surface ${\cal S}$ enclosing volume ${\cal V}$ filled with a homogeneous material  (Fig.~\ref{fig:stratton}), the Stratton-Chu formulation~\cite{Hanson2002} reads
\begin{align}
j\omega \mu_0 &\vect{n} \times \vect{n} \times \Big ( \big [ {\cal L}  \vect{J} \big ] (\vect{r}) + \big[ {\cal L} \left( \vect{n} \times \vect{H} \right) \big](\vect{r}) \Big )  \nonumber \\
&+ \vect{n} \times \vect{n} \times \big[ {\cal K} \left( -\vect{n} \times \vect{E}\right)\big](\vect{r}) = 0 \,.
\label{eq:Stratton-Chu}
\end{align}
Integral operators ${\cal L}$ and ${\cal K}$ in~\eqref{eq:Stratton-Chu} are defined to be~\cite{Gibson2009}
\begin{align}
\left[{\cal L} \vect{X} \right](\vect{r}) &= \left[ 1 + \frac{1}{k_0^2} \nabla \nabla \cdot  \right] \iint_{S_m} G(\vect{r}, \vect{r}') \vect{X}(\vect{r}') d\vect{r}' \,, \\
\left[ {\cal K} \vect{X} \right] (\vect{r}) &= \nabla \times \iint_{S_m} G(\vect{r}, \vect{r}') \vect{X}(\vect{r}') d\vect{r}' \,,
\end{align}
where the wavenumber $k_0 = \omega \sqrt{\mu_0 \varepsilon_0}$ and the Green's function
\begin{equation}
G(\vect{r}, \vect{r}') = \frac{e^{-jk_0 \abs{\vect{r} - \vect{r}'}} }{4\pi \abs{\vect{r} - \vect{r}'}}\,,
\end{equation}
are associated with the material inside ${\cal V}$.
It is important to note that~\eqref{eq:Stratton-Chu} is independent of material and sources outside ${\cal S}$.

\subsection{Stratton-Chu Formulation Applied to the Original Problem}

If the Stratton-Chu formulation is applied to the original problem shown in Fig.~\ref{fig:sample_cell_equivalent_1}, then we obtain \begin{align}
j\omega \mu_0 &\vect{n} \times \vect{n} \times \left( \left[ {\cal L}  \vect{J}_m  \right] (\vect{r}) + \left[ {\cal L} \left( \vect{n} \times \widehat{\vect{H}}_m \right) \right](\vect{r}) \right)  \nonumber \\
&+ \vect{n} \times \vect{n} \times \left[ {\cal K} \left( -\vect{n} \times \widehat{\vect{E}}_m\right)\right](\vect{r}) = 0\,.
\label{eq:EFIE1}
\end{align}
We evaluate this equation twice, first assuming $\vect{r} \in {\cal S}_m$, and then assuming $\vect{r} \in \widehat{\cal S}_m$.
\subsubsection{Surface Integral Equation on ${\cal S}_m$}
\label{sec:SIE_Sm}
We substitute the expansion of fields and currents in~\eqref{eq:J_expand},~\eqref{eq:Hhat_expand}, and~\eqref{eq:Ehat_expand} into~\eqref{eq:EFIE1}.
Next, we test the resulting equation with RWG basis functions $\mathbf{\Lambda}_{m,n}$ on ${\cal S}_m$. 
The resulting system of equations is written compactly as 
\begin{equation}
\mathbb{G}_{m}^{\left(E,J\right)}\mathbb{J}_{m} + \mathbb{G}_{m}^{\left(E, \widehat{H}\right)} \mathbb{\widehat{H}}_{m} +
\mathbb{G}_{m}^{\left(E, \widehat{E}\right)} \mathbb{\widehat{E}}_{m} = 0 \,,
\label{eq:EFIEdiscretized1}
\end{equation}
where entry $(n,n')$ of matrices $\mathbb{G}_{m}^{(E,J)}$, $\mathbb{G}_m^{(E,\widehat{H})}$, and $\mathbb{G}_m^{(E,\widehat{E})}$ is given by
\begin{align}
\left[\mathbb{G}_m^{(E,J)}\right]_{(n,n')} &= j\omega \mu_0 \Big< {\bf \Lambda}_{m,n} (\vect{r}) , \vect{n} \times \vect{n} \times \left[{\cal L} \mathbf{\Lambda}_{m,n'}\right](\vect{r}) \Big > \label{eq:Gm_EJ_1}\\
\left[\mathbb{G}_m^{(E,\widehat{H})}\right]_{(n,n')} &= j\omega \mu_0 \left< {\bf \Lambda}_{m,n}(\vect{r}), \vect{n} \times \vect{n} \times \left[{\cal L} {\bf \widehat{\Lambda}}_{m,n'}\right](\vect{r}) \right >\label{eq:Gm_EJ_2}\\
\left[\mathbb{G}_m^{(E,\widehat{E})}\right]_{(n,n')} &= \left< {\bf \Lambda}_{m,n}(\vect{r}), -\vect{n} \times \vect{n} \times \left[{\cal K} {\bf \widehat{\Lambda}}_{m,n'}'\right](\vect{r}) \right >\label{eq:Gm_EJ_3}\,
\end{align}
and the inner product is defined as
\begin{equation}
\Big < \vect{f}(\vect{r}), \vect{g}(\vect{r}) \Big> = \iint_{S} \vect{f}(\vect{r}) \cdot \vect{g}(\vect{r}) d\vect{r}\,.
\end{equation}

\subsubsection{{Surface Integral Equation on $\widehat{\cal S}_m$}}
Now, we re-evaluate~\eqref{eq:EFIE1} on $\widehat{\cal S}_m$.
We again substitute~\eqref{eq:J_expand},~\eqref{eq:Hhat_expand}, and~\eqref{eq:Ehat_expand} into~\eqref{eq:EFIE1}, and test the resulting equation with the RWG basis functions on $\widehat{\cal S}_m$, i.e.  $\widehat{\mathbf{\Lambda}}_{m,n}$. 
The resulting equations can be compactly written as
\begin{equation}
\mathbb{G}_{m}^{\left(\widehat{E},J\right)}\mathbb{J}_{m} + \mathbb{G}_{m}^{\left(\widehat{E}, \widehat{H}\right)} \mathbb{\widehat{H}}_{m} +
\mathbb{G}_{m}^{\left(\widehat{E}, \widehat{E}\right)} \mathbb{\widehat{E}}_{m} = 0 \,,
\label{eq:EFIEdiscretized2}
\end{equation}
where entry $(n,n')$ of $\mathbb{G}_{m}^{\left(\widehat{E},J\right)}$, $\mathbb{G}_{m}^{\left(\widehat{E}, \widehat{H}\right)}$, and $\mathbb{G}_{m}^{\left(\widehat{E}, \widehat{E}\right)}$ is 
\begin{align}
\left[\mathbb{G}_m^{(\widehat{E},J)}\right]_{(n,n')} &= j\omega \mu_0 \left< \widehat{\bf \Lambda}_{m,n}(\vect{r}), \vect{n} \times \vect{n} \times \left[{\cal L} \mathbf{\Lambda}_{m,n'}\right](\vect{r})\right > \label{eq:G1}\\
\left[\mathbb{G}_m^{(\widehat{E},\widehat{H})}\right]_{(n,n')} &= j\omega \mu_0 \left< \widehat{\bf \Lambda}_{m,n}(\vect{r}), \vect{n} \times \vect{n} \times \left[{\cal L} {\bf \widehat{\Lambda}}_{m,n'}\right](\vect{r})\right >\label{eq:G2} \\
\left[\mathbb{G}_m^{(\widehat{E},\widehat{E})}\right]_{(n,n')} &= \left< \widehat{\bf \Lambda}_{m,n}(\vect{r}), -\vect{n} \times \vect{n} \times \left[{\cal K} {\bf \widehat{\Lambda}}_{m,n'}' \right](\vect{r})\right >\,. \label{eq:G3}
\end{align}
It is important to note that the novel usage of the dual basis functions to expand $\vect{n} \times \widehat{\vect{E}}_m(\vect{r})$ ensures that both ${\cal L}$ and ${\cal K}$ operators in~\eqref{eq:EFIE1} are well-tested~\cite{Sheng1998}. 
Hence, all three discretized matrices in~\eqref{eq:EFIEdiscretized2} are well-conditioned. 
In particular, if we had expanded $\vect{n} \times \widehat{\vect{E}}_m(\vect{r})$ with RWG basis functions, then matrix $\mathbb{G}_m^{(\widehat{E}, \widehat{E})}$ would have been poorly conditioned.

We want to use the discretized Stratton-Chu formulation~\eqref{eq:EFIEdiscretized2} to eliminate $\widehat{\mathbb{H}}_m$ from~\eqref{eq:Jeq2}.
Therefore, since $\mathbb{G}_{m}^{\left(\widehat{E}, \widehat{H}\right)}$ is a well-conditioned matrix, we rewrite~\eqref{eq:EFIEdiscretized2} as
\begin{align}
\mathbb{\widehat{H}}_{m} = - \left[ \mathbb{G}_{m}^{\left(\widehat{E}, \widehat{H}\right)}\right]^{-1}\left[ \mathbb{G}_{m}^{\left(\widehat{E}, \widehat{E}\right)} \mathbb{\widehat{E}}_{m} + \mathbb{G}_{m}^{\left(\widehat{E},J\right)}\mathbb{J}_{m} \right]\,.
\label{eq:Hdiscretized1}
\end{align}
Equation~\eqref{eq:Hdiscretized1} requires the LU factorization of $\mathbb{G}_{m}^{\left(\widehat{E}, \widehat{H}\right)}$, which is a dense matrix. 
However, this matrix is only of size $\widehat{N}_m \times \widehat{N}_m$, and thus the cost of this LU factorization will be relatively small compared to the total time to solve the entire problem.

Equation~\eqref{eq:Hdiscretized1} allows us to eliminate~$\widehat{\mathbb{H}}_m$ from~\eqref{eq:EFIEdiscretized1} and obtain
\begin{align}
&\underbrace{\left[ \mathbb{G}_{m}^{\left(E,J\right)} - \mathbb{G}_{m}^{\left(E, \widehat{H}\right)} \left[ \mathbb{G}_{m}^{\left(\widehat{E}, \widehat{H}\right)} \right]^{-1}  \mathbb{G}_{m}^{\left(\widehat{E},J\right)}\right]}_{\mathbb{A}_m} \mathbb{J}_{m} + \nonumber \\
&\underbrace{\left[ \mathbb{G}_{m}^{\left(E, \widehat{E}\right)} - \mathbb{G}_{m}^{\left(E, \widehat{H}\right)}\left[ \mathbb{G}_{m}^{\left(\widehat{E}, \widehat{H}\right)} \right]^{-1}  \mathbb{G}_{m}^{\left(\widehat{E},\widehat{E}\right)} \right]}_{\mathbb{B}_m} \mathbb{\widehat{E}}_{m} = 0 \,,
\label{eq:JM}
\end{align}
where we have introduced matrices $\mathbb{A}_m$ and $\mathbb{B}_m$. From~\eqref{eq:JM} we have
\begin{align}
{\mathbb{J}}_m = - \mathbb{A}_m^{-1} \mathbb{B}_m \mathbb{\widehat{E}}_m\,,
\label{eq:JABE}
\end{align}
which relates the current density on the PEC scatterers to the tangential electric field on $\widehat{\cal S}_m$.

\subsection{Stratton-Chu Formulation Applied to the Equivalent Problem}

We now apply the Stratton-Chu formulation~\eqref{eq:Stratton-Chu} to the equivalent problem shown in Fig.~\ref{fig:sample_cell_equivalent_2}.
Since there is no electric current distribution inside $\widehat{\cal S}_m$, the Stratton-Chu formulation for $\vect{r} \in \widehat{\cal S}_m$ reads
\begin{align}
j\omega \mu_0 \vect{n} \times \vect{n} \times &\left[ {\cal L} \left( \vect{n} \times \widetilde{\vect{H}}_m \right) \right] (\vect{r})  \nonumber \\ 
&+ \vect{n} \times \vect{n} \times \left[ {\cal K} \left( -\vect{n} \times \widehat{\vect{E}}_m\right)\right] (\vect{r})  = 0 \,,
\label{eq:EFIE3}
\end{align}
which is similar to~\eqref{eq:EFIE1}, except there is no contribution from $\vect{J}_m$.
By testing~\eqref{eq:EFIE3} with RWG basis functions on $\widehat{\cal S}_m$, we obtain
\begin{equation}
\mathbb{G}_{m}^{\left(\widehat{E}, \widehat{H}\right)} \mathbb{\widetilde{H}}_{m} +
\mathbb{G}_{m}^{\left(\widehat{E}, \widehat{E}\right)} \mathbb{\widehat{E}}_{m} = 0 \,,
\label{eq:EFIEdiscretized4}
\end{equation}
where entries of matrices $\mathbb{G}_{m}^{\left(\widehat{E}, \widehat{H}\right)}$ and $\mathbb{G}_{m}^{\left(\widehat{E}, \widehat{E}\right)}$ are given in~\eqref{eq:G2}-\eqref{eq:G3}.
Similarly to~\eqref{eq:Hdiscretized1},~\eqref{eq:EFIEdiscretized4} can be rewritten as 
\begin{equation}
\mathbb{\widetilde{H}}_{m} = - \left[\mathbb{G}_{m}^{\left(\widehat{E}, \widehat{H}\right)} \right]^{-1}
\mathbb{G}_{m}^{\left(\widehat{E}, \widehat{E}\right)} \mathbb{\widehat{E}}_{m} \,,
\label{eq:Hdiscretized2}
\end{equation}
to obtain the tangential magnetic field on $\widehat{\cal S}_m$ in the equivalent problem in terms of the tangential electric field on $\widehat{\cal S}_m$.

\subsection{Equivalent Current}
\label{sec:equivalence_thm}

We can now simplify the expression for the equivalent electric current density on $\widehat{S}_m$ in~\eqref{eq:Jeq2}.
We substitute~\eqref{eq:Hdiscretized1} and~\eqref{eq:Hdiscretized2} into~\eqref{eq:Jeq2} and simplify the resulting equation to obtain
\begin{equation}
\widehat{\mathbb{J}}_{m} = \underbrace{\left[ \mathbb{G}_{m}^{\left(\widehat{E}, \widehat{H}\right)}\right]^{-1}\left[ \mathbb{G}_{m}^{\left(\widehat{E},J\right)} \right]}_{\mathbb{T}_m} \mathbb{J}_{m}\,,
\label{eq:Tm1}
\end{equation}
where $\mathbb{T}_m$ is the transfer matrix that relates the electric current density on ${\cal S}_m$ to the equivalent electric current density on $\widehat{\cal S}_m$.
Since this formulation does not require an equivalent magnetic current density~\cite{AWPL2017}, the proposed formulation is simpler and more efficient than other algorithms in the literature based on the equivalence principle~\cite{li2007}.

\section{Macromodel For Each Element in an Array}
\label{sec:macromodel_multiple}
The macromodel procedure presented in Sec.~\ref{sec:macromodel_single} for a single element is then applied to each element of the array.
We replace the array of scatterers with an array of equivalent electric current densities (equivalent surfaces).
An implicit assumption made here is that none of the array elements touch one another, which is the case in many practical arrays of interest.

The electric current density coefficients on all elements are collected into a vector
\begin{equation}
\mathbb{J} = \begin{bmatrix} {\mathbb{J}}_1^T &  {\mathbb{J}}_1^T & \hdots & {\mathbb{J}}_{M}^T\end{bmatrix}^T
\end{equation}
of size $N \times 1$ where $N = \sum_{m=1}^{M} N_m$. 
Likewise, the coefficients of $\widehat{\vect{J}}_m$ on all equivalent surfaces are collected into a vector
\begin{align}
\widehat{\mathbb{J}} = \begin{bmatrix} \widehat{\mathbb{J}}_1^T &  \widehat{\mathbb{J}}_1^T & \hdots & \widehat{\mathbb{J}}_{M}^T \end{bmatrix}^T \,
\end{align}
of size $\widehat{N} \times 1$ where $\widehat{N} = \sum_{m=1}^{M} \widehat{N}_m$. 
The two current densities, as presented in~\eqref{eq:Tm1}, are related by
\begin{equation}
\widehat{\mathbb{J}} = \mathbb{T} \mathbb{J} \,,
\label{eq:Toperator}
\end{equation}
where
\begin{equation}
\mathbb{T} = 
\begin{bmatrix}
\mathbb{T}_1 & & & \\
& \mathbb{T}_2 & & \\
& & \ddots & \\
& & & \mathbb{T}_M
\end{bmatrix}
\label{eq:Tmatrix}
\end{equation}
is a block-diagonal transfer matrix that relates the equivalent electric current density on $\widehat{{\cal S}}_m$ to the current density on ${\cal S}_m$ for all elements. 
Similar to~\eqref{eq:Tmatrix}, we introduce matrices $\mathbb{A}$ and $\mathbb{B}$ which are block diagonal matrices made up of blocks $\mathbb{A}_m$ and $\mathbb{B}_m$, respectively.
In most array problems, many elements are identical. 
Therefore, matrices $\mathbb{T}_m$, $\mathbb{A}_m$, and $\mathbb{B}_m$ only need to be calculated once for each distinct element.

\section{Exterior Problem}
\label{sec:exterior}
After applying the macromodeling technique, we have simplified the original problem to an equivalent problem composed of an array of equivalent electric current densities.
We apply the electric field integral equation to capture the coupling between the macromodels.

According to the electric field integral equation, the total tangential electric field on the $m$-th equivalent surface is 
\begin{align}
\vect{n} \times \vect{n} \times \widehat{\vect{E}}_m(\vect{r}) = - j\omega\mu_0 \sum_{m'=1}^{M}  &\vect{n} \times \vect{n} \times \left [ {\cal L} \widehat{\vect{J}}_{m'}\right] (\vect{r}) \nonumber \\
&+ \vect{n} \times \vect{n} \times \vect{E}^{(inc)}(\vect{r})\,,
\label{eq:EFIE_Outer}
\end{align}
where the right hand side is the sum of the total scattered field produced by the equivalent electric currents $\widehat{\vect{J}}_{m'}(\vect{r}')$ and the incident electric field $\vect{E}^{(inc)}(\vect{r})$.

We substitute~\eqref{eq:Ehat_expand} and~\eqref{eq:Jeq1} into~\eqref{eq:EFIE_Outer} and test the resulting equation with RWG basis functions $\widehat{\vect{\Lambda}}_{m,n}$ for $m = 1, \hdots, M$. 
The resulting equations can be compactly written as
\begin{equation}
\mathbb{D} \widehat{\mathbb{E}} = -\mathbb{G}^{(\widehat{E}, \widehat{J})} \widehat{\mathbb{J}} + \widehat{\mathbb{V}} \,,
\label{eq:EFIEDiscretized_Outer}
\end{equation}
where $\widehat{\mathbb{E}}$ and $\widehat{\mathbb{V}}$ are vectors of electric field coefficients and the excitation vector, respectively, and read
\begin{align}
\widehat{\mathbb{E}} = \begin{bmatrix}
\widehat{\mathbb{E}}_{1}^T & \widehat{\mathbb{E}}_2^T & \hdots & \widehat{\mathbb{E}}_M^T
\end{bmatrix}^T\,, \\
\widehat{\mathbb{V}} = \begin{bmatrix}
\widehat{\mathbb{V}}_{1}^T & \widehat{\mathbb{V}}_2^T & \hdots & \widehat{\mathbb{V}}_M^T
\end{bmatrix}^T\,. \label{eq:V1}
\end{align}
Vector $\widehat{\mathbb{V}}_m$ in~\eqref{eq:V1} is a vector of size $\widehat{N}_m \times 1$ whose $n$-th entry is
\begin{equation}
\widehat{\mathbb{V}}_{m,n} = \left< {\bf \Lambda}_{m,n}(\vect{r}), \vect{n} \times \vect{n} \times \vect{E}^{(inc)}(\vect{r}) \right>\,,
\end{equation}
which is the projection of incident electric field on RWG basis functions.
In~\pref{eq:EFIEDiscretized_Outer}, $\mathbb{G}^{(\widehat{E}, \widehat{J})}$ is a block matrix of the form
\begin{equation}
\mathbb{G} = \begin{bmatrix}  \mathbb{G}^{\widehat{E}, \widehat{J}}_{1,1} & \mathbb{G}^{\widehat{E}, \widehat{J}}_{1,2} & \hdots & \mathbb{G}^{\widehat{E}, \widehat{J}}_{1,M} \\
\mathbb{G}^{\widehat{E}, \widehat{J}}_{2,1} & \mathbb{G}^{\widehat{E}, \widehat{J}}_{2,2} & \hdots & \mathbb{G}^{\widehat{E}, \widehat{J}}_{2,M} \\
\vdots & \vdots & \ddots & \vdots \\
\mathbb{G}^{\widehat{E}, \widehat{J}}_{M,1} & \mathbb{G}^{\widehat{E}, \widehat{J}}_{M,2} & \hdots & \mathbb{G}^{\widehat{E}, \widehat{J}}_{M,M}
\end{bmatrix}\,,
\end{equation}
where the $(n,n')$ entry is
\begin{equation}
\left[\mathbb{G}^{\widehat{E}, \widehat{J}}_{m,m'}\right]_{n,n'}  = j\omega \mu_0 \left< \widehat{\mathbf{\Lambda}}_{m,n}(\vect{r}), \vect{n} \times \vect{n} \times \left[{\cal L} \widehat{\bf \Lambda}_{m',n'} \right] (\vect{r}) \right>\,.
\label{eq:Gouter1}
\end{equation}
Finally, matrix ${\mathbb{D}}$ in~\eqref{eq:EFIEDiscretized_Outer} is block diagonal and reads
\begin{equation}
{\mathbb{D}} = \begin{bmatrix} 
{\mathbb{D}}_{1} & & \\
& \ddots & \\
& & {\mathbb{D}}_{M}
\end{bmatrix}\,,
\end{equation}
where the $(n,n')$ entry of ${\mathbb{D}}_{m}$ is given by
\begin{equation}
\left[ {\mathbb{D}}_{m}\right]_{(n,n')} = \left< \widehat{\bf \Lambda}_{m,n}(\vect{r}), \vect{n} \times \widehat{\bf \Lambda}_{m,n'}'(\vect{r}) \right>\,.
\end{equation}
It is important to note that ${\mathbb{D}}$ is well-conditioned because it is diagonally dominant, since $\vect{\Lambda}_{m,n}'(\vect{r})$ is approximately orthogonal to $\vect{\Lambda}_{m,n}(\vect{r})$~\cite{Tong2009}.

Next, we subsitute~\eqref{eq:Toperator} and~\eqref{eq:JABE} into the outer problem~\eqref{eq:EFIEDiscretized_Outer} to obtain the final system of equations
\begin{equation}
\left( \mathbb{D} - \mathbb{G}^{(\widehat{E}, \widehat{J})} \mathbb{T} \mathbb{A}^{-1} \mathbb{B} \right) \widehat{\mathbb{E}}  = \widehat{\mathbb{V}}\,,
\label{eq:finalsystem}
\end{equation}
which is only in terms of the electric field coefficients on the surface of the equivalent box.
After solving for $\widehat{\mathbb{E}}$, the current distribution $\mathbb{J}$ on the original scatterer, if desired, can be computed very inexpensively using~\eqref{eq:JABE} for each element.

Notice that solving the original problem with the standard MoM would have required solving for $N$ unknowns. 
However,~\eqref{eq:finalsystem} involves only $\widehat{N}$ unknowns.
For many problems with complex, multiscale scatterers, $\widehat{N}$ is much smaller than $N$, and therefore the proposed method results in faster solution times and lower memory consumption.
Furthermore,~\eqref{eq:finalsystem} is usually better conditioned than the standard MoM formulation because the equivalent surface can have a coarser mesh than the original scatterer due to the absense of any fine features in the equivalent problem.

\section{Acceleration with AIM}
\label{sec:aim}

Even after the reduction in the number of unknowns, the computation cost of the proposed approach can grow quickly.
Therefore, we solve the linear system~\eqref{eq:finalsystem} iteratively using the generalized minimal residual (GMRES) algorithm~\cite{Saad1986}.
This algorithm requires an efficient way to compute
\begin{equation}
\left( \mathbb{D} - \mathbb{G}^{(\widehat{E}, \widehat{J})} \mathbb{T} \mathbb{A}^{-1} \mathbb{B} \right) \mathbb{Y}  \,,
\label{eq:MV_product1}
\end{equation}
where $\mathbb{Y}$ is an arbitrary vector of size $\widehat{N} \times 1$. 
A trivial way to compute~\eqref{eq:MV_product1} is to first generate $\mathbb{D}$, $\mathbb{G}^{(\widehat{E}, \widehat{J})}$, $\mathbb{T}$, $\mathbb{A}$, and $\mathbb{B}$, and then carry out the required matrix-vector multiplications and vector additions.
Since matrices $\mathbb{T}$, $\mathbb{A}$, and $\mathbb{B}$ are block-diagonal matrices with $M$ blocks, computing matrix-vector products with these matrices is inexpensive.
Matrix $\mathbb{D}$ in~\eqref{eq:finalsystem} is also a block-diagonal matrix with $M$ sparse blocks, and so matrix-vector products with this matrix is also inexpensive.
However, $\mathbb{G}^{(\widehat{E}, \widehat{J})}$ is a dense matrix, and so generating this matrix and storing its values requires a lot of computational time and memory.

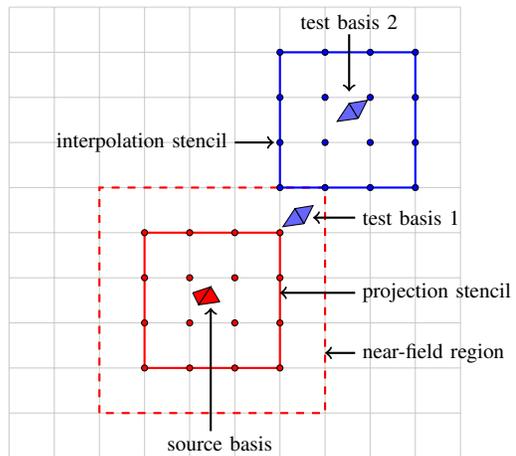
\begin{figure}[t]
\centering
\begin{tikzpicture}[scale=0.8, every node/.style={scale=0.8}]
\draw[step=0.75,black!20,thin] (-0.75,-0.75) grid (6.75,6.75);

\coordinate (p1) at (2.3,2);
\coordinate (p2) at (2.6,2.1);
\coordinate (p3) at (2.4,1.8);

\draw [fill=red] (p1) -- (p2) -- (p3) -- (p1);

\coordinate (p11) at (2.75,1.85);
\coordinate (p12) at (2.6,2.1);
\coordinate (p13) at (2.4,1.8);
\draw [fill=red] (p11) -- (p12) -- (p13) -- (p11);

\draw [red, thick] (1.5,0.75) --++(0,2.25) --++ (2.25,0) --++ (0,-2.25) --++ (-2.25,0);
\draw [red, thick, dashed] (0.75,0.0) --++(0,3.75) --++ (3.75,0) --++ (0,-3.75) --++ (-3.75,0);
\draw [fill=red] (1.5,0.75) circle (0.05);
\draw [fill=red] (2.25,0.75) circle (0.05);
\draw [fill=red] (3.0,0.75) circle (0.05);
\draw [fill=red] (3.75,0.75) circle (0.05);
\draw [fill=red] (1.5,1.5) circle (0.05);
\draw [fill=red] (2.25,1.5) circle (0.05);
\draw [fill=red] (3.0,1.5) circle (0.05);
\draw [fill=red] (3.75,1.5) circle (0.05);
\draw [fill=red] (1.5,2.25) circle (0.05);
\draw [fill=red] (2.25,2.25) circle (0.05);
\draw [fill=red] (3.0,2.25) circle (0.05);
\draw [fill=red] (3.75,2.25) circle (0.05);
\draw [fill=red] (1.5,3.0) circle (0.05);
\draw [fill=red] (2.25,3.0) circle (0.05);
\draw [fill=red] (3.0,3.0) circle (0.05);
\draw [fill=red] (3.75,3.0) circle (0.05);

\coordinate (p11) at (4.15,3.15);
\coordinate (p12) at (4.0,3.4);
\coordinate (p13) at (3.8,3.1);
\draw [fill=blue!60] (p11) -- (p12) -- (p13) -- (p11);

\coordinate (p11) at (4.15,3.15);
\coordinate (p12) at (4.0,3.4);
\coordinate (p13) at (4.3,3.45);
\draw [fill=blue!60] (p11) -- (p12) -- (p13) -- (p11);

\begin{scope}[shift ={(0.9,1.75)}]
\coordinate (p11) at (4.15,3.15);
\coordinate (p12) at (4.0,3.4);
\coordinate (p13) at (3.8,3.1);
\draw [fill=blue!60] (p11) -- (p12) -- (p13) -- (p11);

\coordinate (p11) at (4.15,3.15);
\coordinate (p12) at (4.0,3.4);
\coordinate (p13) at (4.3,3.45);
\draw [fill=blue!60] (p11) -- (p12) -- (p13) -- (p11);
\end{scope}

\draw [blue, thick] (3.75,3.75) --++(0,2.25) --++ (2.25,0) --++ (0,-2.25) --++ (-2.25,0);

\draw [fill=blue] (3.75,3.75) circle (0.05);
\draw [fill=blue] (4.5,3.75) circle (0.05);
\draw [fill=blue] (5.25,3.75) circle (0.05);
\draw [fill=blue] (6.0,3.75) circle (0.05);
\draw [fill=blue] (3.75,4.5) circle (0.05);
\draw [fill=blue] (4.5,4.5) circle (0.05);
\draw [fill=blue] (5.25,4.5) circle (0.05);
\draw [fill=blue] (6.0,4.5) circle (0.05);
\draw [fill=blue] (3.75,5.25) circle (0.05);
\draw [fill=blue] (4.5,5.25) circle (0.05);
\draw [fill=blue] (5.25,5.25) circle (0.05);
\draw [fill=blue] (6.0,5.25) circle (0.05);
\draw [fill=blue] (3.75,6.0) circle (0.05);
\draw [fill=blue] (4.5,6.0) circle (0.05);
\draw [fill=blue] (5.25,6.0) circle (0.05);
\draw [fill=blue] (6.0,6.0) circle (0.05);

\node at (5,2) [right] {projection stencil};
\draw [->, thick] (5,2) -- (3.75,2);
\node at (5,1) [right] {near-field region};
\draw [->, thick] (5,1) -- (4.5,1);

\node at (5,3.25) [right] {test basis 1};
\draw [->, thick] (5,3.25) -- (4.3,3.25);

\node at (2.75,-0.5) {source basis};
\draw [->, thick] (2.6,-.3) -- (2.60,1.75);

\node at (3,4.5) [left] {interpolation stencil};
\draw [->, thick] (3,4.5) -- (3.65,4.5);

\node at (4.9,6.5) {test basis 2};
\draw [->, thick] (4.9,6.3) -- (4.9,5.35);

\end{tikzpicture}
\caption{Sample AIM grid with one source RWG basis function  and two testing functions. The inner product between the source function and test function 1 is calculated directly. The inner product between source function and test function 2 is computed via AIM.}
\label{fig:aim}
\end{figure}

For this reason, we accelerate the computation of $\mathbb{G}^{(\widehat{E}, \widehat{J})} \mathbb{Y}$ using 
AIM~\cite{Bleszynski96,Zhu2005}. In AIM, the problem domain is discretized using a 3D Cartesian grid.
Each basis function is assigned to a stencil. 
Each stencil is made up of $N_{O}+1$ grid points in each direction, where $N_O$ is the stencil order. 
Furthermore, we define the $N_{NF}$ stencils surrounding the basis function in each direction to be the near-field region associated with each basis function.
Figure~\ref{fig:aim} shows a sample 2D AIM grid with $N_O = 3$ and $N_{NF} = 1$.

For AIM, $\mathbb{G}^{(\widehat{E}, \widehat{J})}$ is decomposed into two parts: the near-field matrix $\mathbb{G}^{(\widehat{E}, \widehat{J})}_{NF} $ and the far-field matrix $\mathbb{G}^{(\widehat{E}, \widehat{J})}_{FF}$.
The near-field matrix is a sparse matrix whose elements are computed by~\eqref{eq:Gouter1} with some pre-correction~\cite{Bleszynski96,Zhu2005}.
The far-field matrix, on the other hand, is factorized as a product of three matrices as
\begin{equation}
\mathbb{G}^{(\widehat{E}, \widehat{J})}_{FF} = \sum_{\psi = \{A_x, A_y, A_z, \phi\}} \mathbb{I}_{\psi} \mathbb{H}_{\psi} \mathbb{P}_{\psi}\,,
\end{equation}
where $\mathbb{I}_{\psi}$, $\mathbb{H}_{\psi}$, and $\mathbb{P}_{\psi}$ are the interpolation, convolution, and projection matrices, respectively. 
These matrices, in general, need to be calculated for the scalar potential $\phi$ and vector potentials in the three principal directions $x$, $y$, and $z$.

Given $\mathbb{Y}$, the matrix-vector product $\mathbb{G}^{(\widehat{E}, \widehat{J})}_{FF} \mathbb{Y}$ is computed in AIM by the following steps:
\begin{enumerate}[(i)]
\item Compute equivalent grid charges and grid currents (in the $x$, $y$, and $z$ directions) on the projection stencil using interpolation polynomials of order $N_O$. These equivalent grid charges and currents produce the same fields as the original source basis functions in the \emph{far field}. Mathematically, this operation can be expressed as
\begin{equation}
\mathbb{Y}^{(1)}_{\psi} =\mathbb{P}_{\psi} \mathbb{Y} \,,
\end{equation} 
where $\mathbb{P}_{\psi}$ is a sparse matrix.
\item Compute grid scalar and vector potentials from the grid charges and currents. Mathematically, this operation can be expressed as
\begin{equation}
\mathbb{Y}_{\psi}^{(2)} = \mathbb{H}_{\psi} \mathbb{Y}^{(1)}_{\psi} \,.
\label{eq:AIM_step2}
\end{equation}
$\mathbb{H}_{\psi}$ is a dense matrix with size equal to the number of grid points, hence storing $\mathbb{H}_{\psi}$ can be very expensive. 
However, due to the position-invariance property of the Green's function, the fast Fourier transform (FFT) is applied to compute the matrix-vector product in~\eqref{eq:AIM_step2}.
\item From the grid potentials calculated in step (ii), compute the electric field and its projection on the testing  functions. To find the electric field on the basis function from the grid potentials, we use an interpolation polynomial of order $N_O$. Mathematically, this operation can be expressed as
\begin{equation}
\mathbb{Y}^{(3)}_{\psi} = \mathbb{I}_{\psi} \mathbb{Y}^{(2)}_{\psi}\,,
\end{equation}
where $\psi \in \left \{ \phi, A_x, A_y,A_z \right \}$. The interpolation matrix $\mathbb{I}_{\psi}$ is a sparse matrix.
\end{enumerate}
Finally, the result of matrix-vector product is $\mathbb{G}_{FF}^{(\widehat{E}, \widehat{J})} \mathbb{Y} = \sum_{\psi = \{A_x, A_y, A_z, \phi\}} \mathbb{Y}^{(3)}_{\psi}$.
Readers interested to learn more about AIM are referred to~\cite{Bleszynski96,Zhu2005}.

\section{Numerical Results}
\label{sec:Results}

In this section, three examples are presented to demonstrate the accuracy and performance of the proposed method compared against an in-house standard MoM solver accelerated with AIM.
All computational codes were developed using PETSc~\cite{petsc-web-page,petsc-user-ref,petsc-efficient} and FFTW3~\cite{FFTW05} libraries.
The numerical tests were performed on a machine with an Intel Xeon E5-2623 v3 processor and 128~GB of RAM.
All simulations were run on a single thread without exploiting any parallelization.

\subsection{Array of Spherical Helix Antennas}
\label{sec:helix}

\begin{figure}[t]
\centering
\input{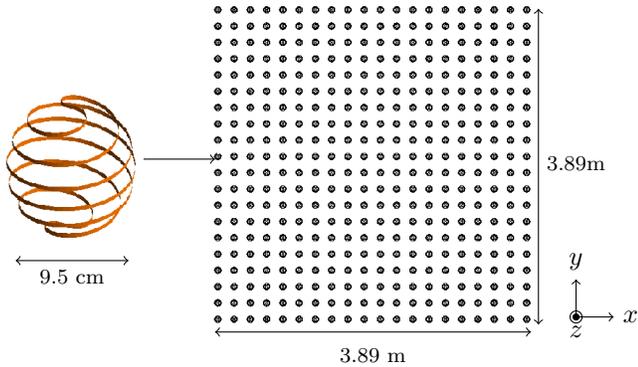}
\caption{Layout of the $20 \times 20$ array of spherical helix antennas considered in Sec.~\ref{sec:helix}. The array is uniformly spaced along the $x-$ and $y$-directions with interelement spacing  of $d_x = d_y = 0.2~{\rm m}$.}
\label{fig:helixfig}
\end{figure}

\begin{figure}
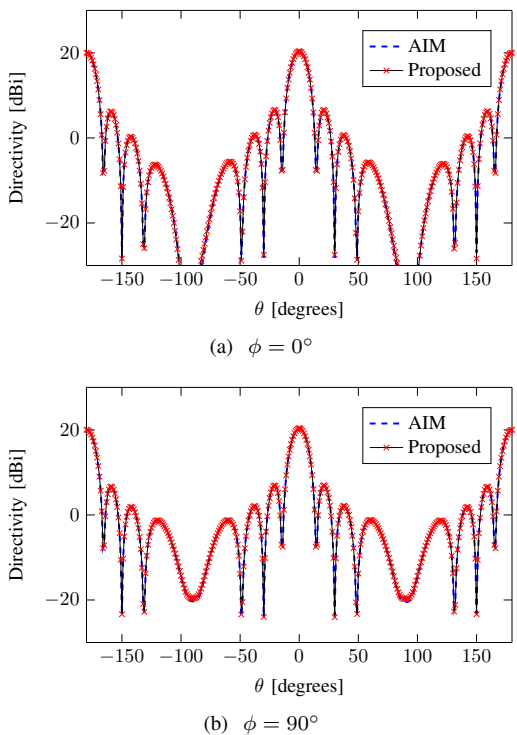

\begin{center}
\subfloat[\label{fig:reflectarray_cut1} $\phi = 0^{\circ}$] {\resizebox {0.8\columnwidth} {!} {
\input{./Results/SphericalHelix/cut1.tex}
}}\\
\subfloat[\label{fig:reflectarray_cut2} $\phi = 90^{\circ}$] {\resizebox {0.8\columnwidth} {!} {
\input{./Results/SphericalHelix/cut2.tex}
}}\\
\end{center}
\caption{Directivity of the $20 \times 20$ array of spherical helix antennas considered in Sec.~\ref{sec:helix} calculated with AIM and the proposed technique.}
\label{fig:directivity_helix}
\end{figure}

\begin{table}[t]
\caption{Simulation Settings and Results for the $20 \times 20$ array of spherical helix antennas considered in Sec.~\ref{sec:helix}}
\label{tab:helix}
\begin{center}
\begin{tabular}{|l| c | c|}
\hline
& AIM & Proposed \\
\hline
\multicolumn{3}{|c|}{AIM Parameters}\\ \hline
Number of stencils & \multicolumn{2}{c|}{$120 \times 120 \times 2$} \\
Interpolation order & \multicolumn{2}{c|}{3} \\ 
Number of near-field stencils & \multicolumn{2}{c|}{4} \\
\hline
\multicolumn{3}{|c|}{Memory Consumption}\\
\hline
Total number of unknowns & 260,800 & 62,400 \\
Memory used & 19.53~GB & 3.60~GB \\
\hline
\multicolumn{3}{|c|}{Timing Results}\\
\hline
Macromodel generation & N/A & 18~s \\
Matrix fill time & 1.12~h &  336~s \\
Preconditioner factorization & 414~s & 23~s \\
Iterative solver & 28~s & 4~s \\
Total computation time & 1.25~h & 6.35~min \\
\hline
\end{tabular}
\end{center}
\end{table}

Spherical helix antennas, like many other electrically small antennas, have complex geometries with electrically fine features. 
Therefore, simulating an array of such antennas requires a long computation time and large memory. 
In the proposed technique, a macromodel is created for each small antenna. 
As demonstrated by this example, the macromodel can accurately capture radiation from the antenna using fewer  unknowns, which leads to significant savings in computation time and memory.
This example also demonstrates that the proposed macromodel approach can be applied to fed antenna arrays.

We consider the $20 \times 20$ array of identical spherical helix antennas shown in Fig.~\ref{fig:helixfig}.
Each element of the array is excited with a uniform delta-gap voltage source at the center of each helix operating at $300~\mathrm{MHz}$. With this excitation, the array radiates with main beam in the broadside direction.
We computed the radiation pattern from the antenna array using the standard MoM and the proposed method, both accelerated with AIM and solved iteratively using GMRES with an ILU-2 preconditioner~\cite{petsc-user-ref}.
In the proposed method, we first created the macromodel for a spherical helix antenna by computing $\mathbb{T}_m$, $\mathbb{A}_m$, and $\mathbb{B}_m$ using a sphere of radius $5~{\rm cm}$ as the equivalent surface. 
This macromodel was then reused for all elements of the array, as described in Sec.~\ref{sec:macromodel_multiple}. 
The AIM parameters, computational times, and memory requirements to simulate this problem are given in Tab.~\ref{tab:helix}.
As seen from Tab.~\ref{tab:helix}, the proposed method reduces the number of unknowns by a factor of 4. 
This reduction in the number of unknowns leads to a simulation that is $12$ times faster and requires $5$ times less memory than the AIM-accelerated MoM solver.
As evident from Tab.~\ref{tab:helix}, the proposed method is faster than AIM because it takes less time to assemble  all the matrices and factorize the preconditioner, since it has fewer unknowns than the standard MoM formulation.
Figure~\ref{fig:directivity_helix} shows the radiation pattern of the array in the two principal plane cuts.
Results in Fig.~\ref{fig:directivity_helix} confirm that the proposed macromodeling approach provides an excellent accuracy compared to the standard MoM code.

\subsection{Two-layer Reflectarray with Jerusalem Cross Elements}
\label{sec:jerusalemcross}

\begin{figure}[t]
\null \hfill
\subfloat[\label{fig:reflectarray_unitcell_original} Original] {\begin{tikzpicture}[xscale= 0.4]
\begin{scope}[every node/.append style={xslant=3},xslant=3, xscale= 1, yscale = 0.15]
\coordinate (p1) at (-\UnitcellWidth*0.5,-\UnitcellLength*0.5);
\coordinate (p2) at (\UnitcellWidth*0.5,-\UnitcellLength*0.5);
\coordinate (p3) at (\UnitcellWidth*0.5, \UnitcellLength*0.5);
\coordinate (p4) at (-\UnitcellWidth*0.5,\UnitcellLength*0.5);
\draw[fill=brown!90] (p1)--(p2)--(p3)--(p4)--(p1);
\end{scope}
\node at (-\UnitcellWidth*0.1,-\UnitcellLength*0.15) {\footnotesize $w$};
\node at (\UnitcellWidth*0.7,-\UnitcellLength*0.02) {\footnotesize $w$};
\draw[<->] (-\UnitcellWidth*0.68, -\UnitcellWidth*0.1) --(\UnitcellWidth*0.28, -\UnitcellWidth*0.1);

\draw[<->] (\UnitcellWidth*0.80, \UnitcellWidth*0.078) --(\UnitcellWidth*0.31, -\UnitcellWidth*0.085);

\draw[<->](\UnitcellWidth*0.80, \h*0.4) -- (\UnitcellWidth*0.8, \h*1.3);

\draw[<->](\UnitcellWidth*0.80, \h*1.4) -- (\UnitcellWidth*0.8, \h*2.3);

\node at (\UnitcellWidth*0.8, \h*0.85) [right] {\footnotesize $h$};
\node at (\UnitcellWidth*0.8, \h*1.85) [right] {\footnotesize $h$};

\draw[fill=blue!40!white, opacity=0.3] (p1) -- (p2) --++(0,\h) --++(-\UnitcellWidth,0) -- (p1);
\begin{scope}[shift = {($(0,\h)$)}, every node/.append style={xslant=3},xslant=3, xscale= 1, yscale = 0.15]

\coordinate (p1) at (-\UnitcellWidth*0.5,-\UnitcellLength*0.5);
\coordinate (p2) at (\UnitcellWidth*0.5,-\UnitcellLength*0.5);
\coordinate (p3) at (\UnitcellWidth*0.5, \UnitcellLength*0.5);
\coordinate (p4) at (-\UnitcellWidth*0.5,\UnitcellLength*0.5);
\draw[fill=blue!40!white, opacity=0.3] (p1)--(p2)--(p3)--(p4)--(p1);

\coordinate (pp+1) at ($(-0.5*\lcxl, 0.5*\ldyl )$);
\coordinate (pp+2) at ($(\lcxl*0.5, \ldyl*0.5)$);
\coordinate (pp+3) at ($(\lcxl*0.5, \ldyl*0.5 - \wcyl)$);
\coordinate (pp+4) at ($(\wdxl*0.5, \ldyl*0.5 - \wcyl)$);
\coordinate (pp+5) at ($(\wdxl*0.5, \wdyl*0.5)$);
\coordinate (pp+6) at ($(\ldxl*0.5 - \wcxl, \wdyl*0.5)$);
\coordinate (pp+7) at ($(\ldxl*0.5 - \wcxl, \lcyl*0.5)$);
\coordinate (pp+8) at ($(\ldxl*0.5, \lcyl*0.5)$);
\coordinate (pp+9) at ($(\ldxl*0.5, -\lcyl*0.5)$);
\coordinate (pp+10) at ($(\ldxl*0.5- \wcxl, -\lcyl*0.5)$);
\coordinate (pp+11) at ($(\ldxl*0.5 - \wcxl,  -\wdyl*0.5)$);
\coordinate (pp+12) at ($(\wdxl*0.5, -\wdyl*0.5)$);
\coordinate (pp+13) at ($(\wdxl*0.5, -\ldyl*0.5 + \wcyl)$);
\coordinate (pp+14) at ($(\lcxl*0.5,  -\ldyl*0.5 + \wcyl)$);
\coordinate (pp+15) at ($(\lcxl*0.5, -\ldyl*0.5)$);
\coordinate (pp+16) at ($(-\lcxl*0.5, -\ldyl*0.5)$);
\coordinate (pp+17) at ($(-\lcxl*0.5, -\ldyl*0.5+ \wcyl)$);
\coordinate (pp+18) at ($(-\wdxl*0.5, -\ldyl*0.5+\wcyl)$);
\coordinate (pp+19) at ($(-\wdxl*0.5,  -\wdyl*0.5)$);
\coordinate (pp+20) at ($(-\ldxl*0.5 + \wcxl, -\wdyl*0.5)$);
\coordinate (pp+21) at ($(-\ldxl*0.5 + \wcxl,  -\lcyl*0.5)$);
\coordinate (pp+22) at ($(-\ldxl*0.5,  -\lcyl*0.5)$);
\coordinate (pp+23) at ($(-\ldxl*0.5, \lcyl*0.5)$);
\coordinate (pp+24) at ($(-\ldxl*0.5+\wcxl, \lcyl*0.5)$);
\coordinate (pp+25) at ($(-\ldxl*0.5+\wcxl,  \wdyl*0.5)$);
\coordinate (pp+26) at ($(-\wdxl*0.5,  \wdyl*0.5)$);
\coordinate (pp+27) at ($(-\wdxl*0.5,  \ldyl*0.5-\wcyl)$);
\coordinate (pp+28) at ($(-\lcxl*0.5,  \ldyl*0.5-\wcyl)$);

\draw[fill = brown!70] (pp+1) -- (pp+2) -- (pp+3) -- (pp+4) -- (pp+5) -- (pp+6) -- (pp+7) -- (pp+8) -- (pp+9) --(pp+10) --(pp+11) --(pp+12) --(pp+13) --(pp+14) --(pp+15) --(pp+16) --(pp+17) --(pp+18) --(pp+19) --(pp+20) --(pp+21) --(pp+22) --(pp+22) --(pp+23) --(pp+24) --(pp+25) --(pp+26) --(pp+27) --(pp+28) --(pp+1);
\end{scope}
\draw[fill=blue!40!white, opacity=0.3] (p2) -- (p3) --++(0,-\h) -- ($(p2) + (0,-\h)$) -- (p2);

\draw[fill=blue!40!white, opacity=0.3] (p1) -- (p2) --++(0,\h) --++(-\UnitcellWidth,0) -- (p1);
\begin{scope}[shift = {($(0,2*\h)$)}, every node/.append style={xslant=3},xslant=3, xscale= 1, yscale = 0.15]

\coordinate (p1) at (-\UnitcellWidth*0.5,-\UnitcellLength*0.5);
\coordinate (p2) at (\UnitcellWidth*0.5,-\UnitcellLength*0.5);
\coordinate (p3) at (\UnitcellWidth*0.5, \UnitcellLength*0.5);
\coordinate (p4) at (-\UnitcellWidth*0.5,\UnitcellLength*0.5);
\draw[fill=blue!40!white, opacity=0.3] (p1)--(p2)--(p3)--(p4)--(p1);

\coordinate (pp+1) at ($(-0.5*\lcxl, 0.5*\ldyl )$);
\coordinate (pp+2) at ($(\lcxl*0.5, \ldyl*0.5)$);
\coordinate (pp+3) at ($(\lcxl*0.5, \ldyl*0.5 - \wcyl)$);
\coordinate (pp+4) at ($(\wdxl*0.5, \ldyl*0.5 - \wcyl)$);
\coordinate (pp+5) at ($(\wdxl*0.5, \wdyl*0.5)$);
\coordinate (pp+6) at ($(\ldxl*0.5 - \wcxl, \wdyl*0.5)$);
\coordinate (pp+7) at ($(\ldxl*0.5 - \wcxl, \lcyl*0.5)$);
\coordinate (pp+8) at ($(\ldxl*0.5, \lcyl*0.5)$);
\coordinate (pp+9) at ($(\ldxl*0.5, -\lcyl*0.5)$);
\coordinate (pp+10) at ($(\ldxl*0.5- \wcxl, -\lcyl*0.5)$);
\coordinate (pp+11) at ($(\ldxl*0.5 - \wcxl,  -\wdyl*0.5)$);
\coordinate (pp+12) at ($(\wdxl*0.5, -\wdyl*0.5)$);
\coordinate (pp+13) at ($(\wdxl*0.5, -\ldyl*0.5 + \wcyl)$);
\coordinate (pp+14) at ($(\lcxl*0.5,  -\ldyl*0.5 + \wcyl)$);
\coordinate (pp+15) at ($(\lcxl*0.5, -\ldyl*0.5)$);
\coordinate (pp+16) at ($(-\lcxl*0.5, -\ldyl*0.5)$);
\coordinate (pp+17) at ($(-\lcxl*0.5, -\ldyl*0.5+ \wcyl)$);
\coordinate (pp+18) at ($(-\wdxl*0.5, -\ldyl*0.5+\wcyl)$);
\coordinate (pp+19) at ($(-\wdxl*0.5,  -\wdyl*0.5)$);
\coordinate (pp+20) at ($(-\ldxl*0.5 + \wcxl, -\wdyl*0.5)$);
\coordinate (pp+21) at ($(-\ldxl*0.5 + \wcxl,  -\lcyl*0.5)$);
\coordinate (pp+22) at ($(-\ldxl*0.5,  -\lcyl*0.5)$);
\coordinate (pp+23) at ($(-\ldxl*0.5, \lcyl*0.5)$);
\coordinate (pp+24) at ($(-\ldxl*0.5+\wcxl, \lcyl*0.5)$);
\coordinate (pp+25) at ($(-\ldxl*0.5+\wcxl,  \wdyl*0.5)$);
\coordinate (pp+26) at ($(-\wdxl*0.5,  \wdyl*0.5)$);
\coordinate (pp+27) at ($(-\wdxl*0.5,  \ldyl*0.5-\wcyl)$);
\coordinate (pp+28) at ($(-\lcxl*0.5,  \ldyl*0.5-\wcyl)$);

\draw[fill = brown!70] (pp+1) -- (pp+2) -- (pp+3) -- (pp+4) -- (pp+5) -- (pp+6) -- (pp+7) -- (pp+8) -- (pp+9) --(pp+10) --(pp+11) --(pp+12) --(pp+13) --(pp+14) --(pp+15) --(pp+16) --(pp+17) --(pp+18) --(pp+19) --(pp+20) --(pp+21) --(pp+22) --(pp+22) --(pp+23) --(pp+24) --(pp+25) --(pp+26) --(pp+27) --(pp+28) --(pp+1);
\end{scope}
\draw[fill=blue!40!white, opacity=0.3] (p2) -- (p3) --++(0,-\h) -- ($(p2) + (0,-\h)$) -- (p2);
\end{tikzpicture}}
\hfill
\subfloat[\label{fig:reflectarray_unitcell_image} Equivalent] {\input{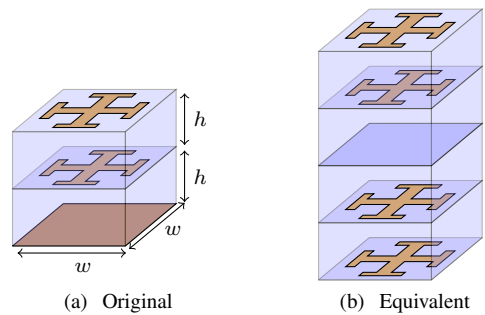}}
\hfill \null
\caption{(a): Original unit cell of the two-layer reflectarray considered in Sec.~\ref{sec:jerusalemcross} with $w=3.75~{\rm mm}$ and $h = 0.76~{\rm mm}$. (b): Equivalent unit cell obtained after applying the image theorem.}
\end{figure}

\begin{figure}[t]
\begin{center}
\input{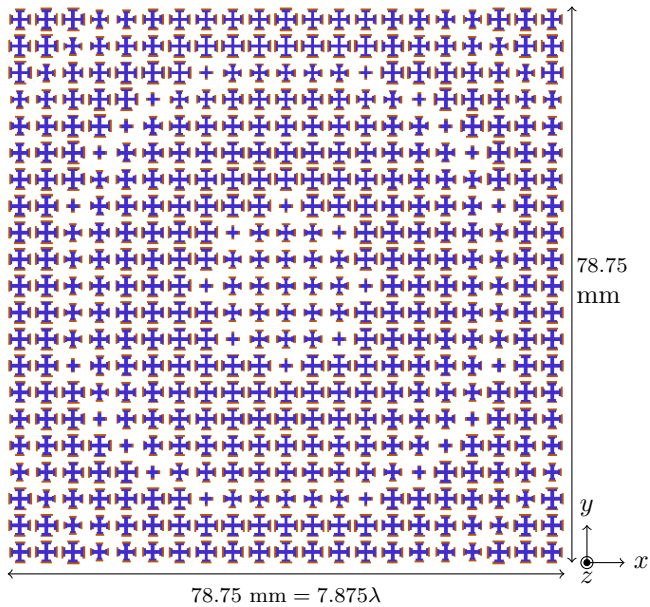}
\end{center}
\caption{Top view of the $21 \times 21$ reflectarray considered in Sec.~\ref{sec:jerusalemcross}. Top and bottom layers of the reflectarray are shown in blue and red, respectively. The reflectarray is uniformly spaced along the $x$- and $y$-directions with interelement spacing of $3.75~{\rm mm}$ ($0.375 \lambda$).  }
\label{fig:TopView_reflectarrayJCross}
\end{figure}

\begin{table}[t]
\caption{Simulation Settings and Results for the $21 \times 21$ reflectarray considered in Sec.~\ref{sec:jerusalemcross}}
\label{tab:reflectarray_Jcross}
\begin{center}
\begin{tabular}{|l| c | c|c|}
\hline
& AIM & ADF & Proposed \\
\hline
\multicolumn{4}{|c|}{AIM Parameters}\\ \hline
Number of stencils in $x$ dir. & $126 $ & - & $84$ \\
Number of stencils in $y$ dir. & $126 $ & - & $84$ \\
Number of stencils in $z$ dir. & $2$ & - & $2$ \\
Interpolation order & 3 & - & 3  \\ 
Number of near-field stencils & 4 & -   & 4 \\
\hline
\multicolumn{4}{|c|}{Memory Consumption}\\
\hline
Total number of unknowns & 324,420 & 324,420 & 111,132\\
Memory used & 40~GB & 37~GB & 15~GB \\
\hline
\multicolumn{4}{|c|}{Timing Results}\\
\hline
Macromodel generation & N/A & N/A & 0.054~h \\
Matrix fill time & 1.48~h &  2.26~h & 0.42~h\\
Preconditioner factorization & 1.25~h & 1.02~h & 0.31~h \\
Iterative solver & 0.22~h & 0.11~h & 2.80~min\\
Total computation time & 3.30~h & 3.40~h & 0.82~h\\
\hline
\end{tabular}
\end{center}
\end{table}

\subsubsection{Design and Simulation Setup}
Next, we consider a two-layer dual-polarized reflectarray with $21 \times 21$ elements made up of Jerusalem crosses~\cite{Geaney2017}. 
The unit cell of the reflectarray is shown in Fig.~\ref{fig:reflectarray_unitcell_original}.
In this example, the reflectarray is electrically large with dimensions of $7.875 \lambda \times 7.875 \lambda$ at $30~{\rm GHz}$.
It also includes sub-wavelength features and strong mutual coupling 
between the unit cells.  
In practice, due to simulation difficulties, reflectarrays of this size and complexity are rarely simulated with full-wave electromagnetic solvers. 
One of the motivations of this work is to enable an efficient simulation of such problems.

Since the proposed method currently does not support multilayer dielectrics, we assume that all layers of the reflectarray have permittivity $\varepsilon_0$ and permeability $\mu_0$. 
Furthermore, we apply the image theory~\cite{Bal05} to model the ground plane at the bottom of the reflectarray. 
According to the image theory,  the two Jerusalem crosses in each unit cell are duplicated below the image plane as shown in Fig.~\ref{fig:reflectarray_unitcell_image}. 
Thus, each unit cell has effectively four Jerusalem crosses.

In this example, the reflectarray is designed to produce the main beam of the scattered field in the broadside direction when the reflectarray is excited by a dipole feed antenna operating at $30~\mathrm{GHz}$\footnote{A dipole feed antenna was chosen due to its simplicity, although it is not an optimal feed model for reflectarrays.}. The feed is  placed $40~\mathrm{mm}$ along the axis of the reflectarray at the prime focus position, so that the focal length to diameter ratio is 0.51. 
The top view of the final reflectarray design is shown in Fig.~\ref{fig:TopView_reflectarrayJCross}.
The final design contains 441 total elements with eight unique elements used to discretize the reflectarray phase curve.

\subsubsection{Scattered Field}

\begin{figure}
\begin{center}
\subfloat[\label{fig:reflectarray_cut1} $\phi = 0^{\circ}$] {\resizebox {0.8\columnwidth} {!} {
\input{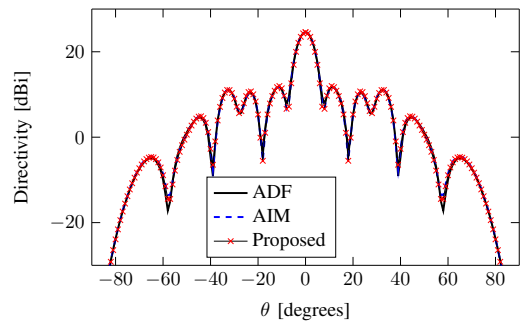}
}}\\
\subfloat[\label{fig:reflectarray_cut2} $\phi = 90^{\circ}$] {\resizebox {0.8\columnwidth} {!} {
\input{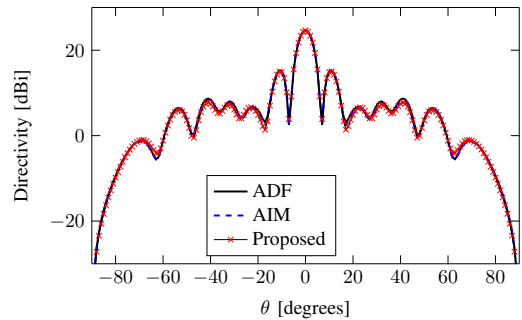}
}}\\
\subfloat[\label{fig:reflectarray_cut3} $\phi = 45^{\circ}$] {\resizebox {0.8\columnwidth} {!} {\input{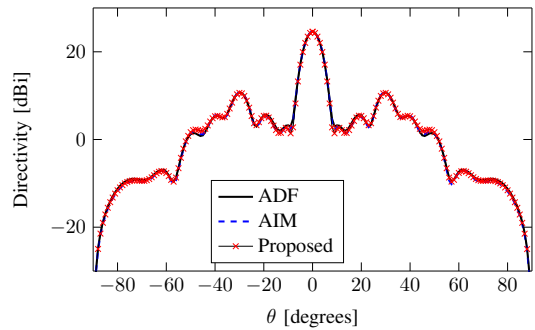}}}
\end{center}
\caption{Directivity of the $21 \times 21$ two-layer reflectarray considered in Sec.~\ref{sec:jerusalemcross} calculated with ADF, AIM, and the proposed technique.}
\label{fig:reflectarray_jerusalemcross}
\end{figure}

We calculated the scattered field from the reflectarray using three tools: an in-house AIM accelerated MoM code, the  Antenna Design Framework (ADF)~\cite{Sabbadini2009} -- an AIM-accelerated commercial SIE solver, and the proposed technique.
For the proposed method, we enclosed each element with an equivalent box of dimensions $ 3.60~{\rm mm} \times 3.60~{\rm mm} \times 4.00~{\rm mm}$.
We generated macromodels for the array by first computing $\mathbb{T}_m$, $\mathbb{A}_m$, and $\mathbb{B}_m$ for the eight unique elements in the array, and then generating $\mathbb{T}$, $\mathbb{A}$, and $\mathbb{B}$.
Figure~\ref{fig:reflectarray_jerusalemcross} shows the directivity in three planes generated with the proposed method, the in-house MoM code, and the ADF solver. 
An excellent match between all three methods validates the accuracy of the proposed method. 
In particular, despite of the unit cells being very close to one another, the macromodel approach accurately predicts the mutual coupling between them.
Simulation settings, memory consumption, and timing results of the simulations run with the in-house MoM code, ADF, and the proposed method are given in Tab.~\ref{tab:reflectarray_Jcross}. 
It is seen that the in-house AIM-accelerated MoM code performs on-par with the commercial AIM-accelerated MoM code.
For this simulation, the proposed method requires $4$ times less computational time and $2.7$ times less memory, which is a substantial savings.

\subsection{Reflectarray Composed of Elements with Fine Features}
\label{sec:meanderreflectarray}

\begin{figure}[t]
\begin{center}
\input{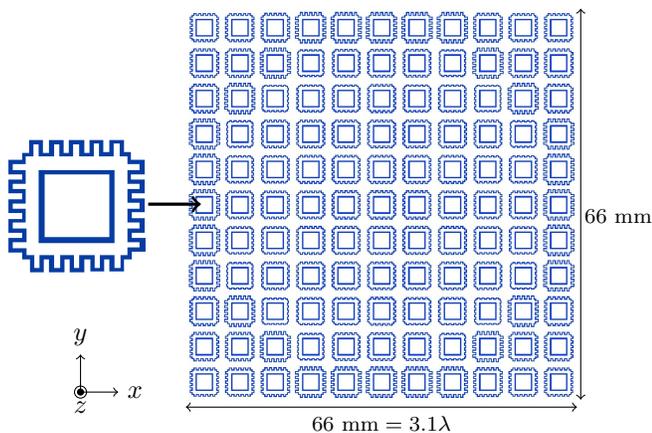}
\end{center}
\caption{Top view of the $11 \times 11$ reflectarray considered in Sec.~\ref{sec:meanderreflectarray}. The array is uniformly spaced along the $x-$ and $y$-directions with interelement spacing of $d_x = d_y = 6~{\rm mm}$ ($0.28~\lambda$)}
\label{fig:reflectarray_meander}
\end{figure}

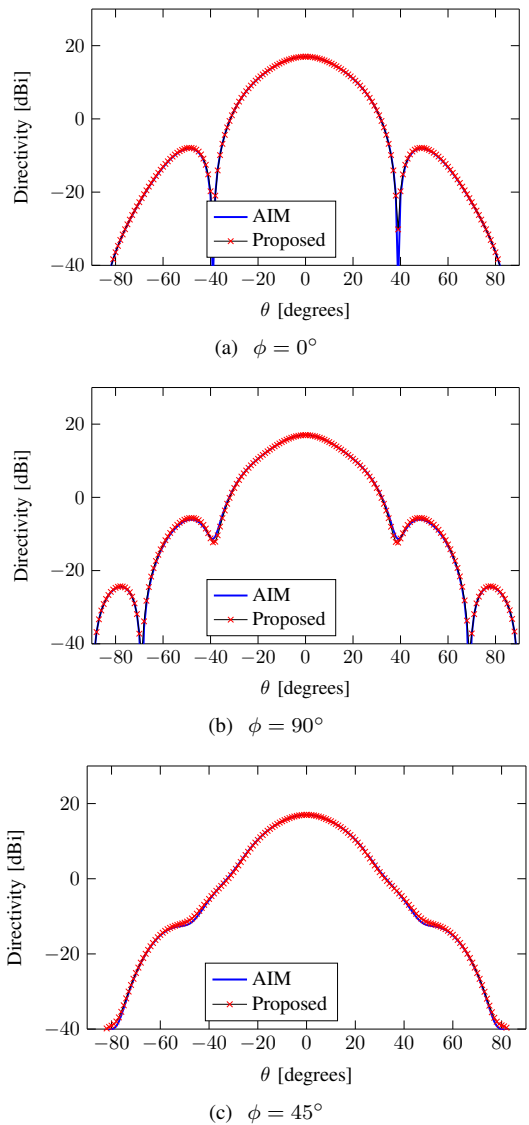
\begin{figure}[t]
\begin{center}
\subfloat[\label{fig:reflectarray_cut1} $\phi = 0^{\circ}$] {\resizebox {0.8\columnwidth} {!} {
\begin{tikzpicture}

\begin{axis}[width=3.0in,
height=1.8in,
scale only axis,
separate axis lines,
every outer x axis line/.append style={black},
every x tick label/.append style={font=\color{black}},
every x tick/.append style={black},
xmin=-90,
xmax=90,
xlabel={$\theta$ [degrees]},
every outer y axis line/.append style={black},
every y tick label/.append style={font=\color{black}},
every y tick/.append style={black},
ymin=-40,
ymax=30,
ylabel={Directivity [dBi]},
axis background/.style={fill=white},
legend style={legend cell align=left, align=left, draw=black,at={(0.42,0.03)},anchor=south}
]

\addplot [color=blue, line width=1pt]
  table[row sep=crcr]{%
  -90.0000  -81.4697\\
  -89.0000  -75.4325\\
  -88.0000  -64.3817\\
  -87.0000  -57.4339\\
  -86.0000  -52.4659\\
  -85.0000  -48.5990\\
  -84.0000  -45.4295\\
  -83.0000  -42.7395\\
  -82.0000  -40.3974\\
  -81.0000  -38.3175\\
  -80.0000  -36.4405\\
  -79.0000  -34.7232\\
  -78.0000  -33.1332\\
  -77.0000  -31.6457\\
  -76.0000  -30.2412\\
  -75.0000  -28.9039\\
  -74.0000  -27.6220\\
  -73.0000  -26.3860\\
  -72.0000  -25.1886\\
  -71.0000  -24.0249\\
  -70.0000  -22.8914\\
  -69.0000  -21.7862\\
  -68.0000  -20.7084\\
  -67.0000  -19.6582\\
  -66.0000  -18.6366\\
  -65.0000  -17.6450\\
  -64.0000  -16.6852\\
  -63.0000  -15.7596\\
  -62.0000  -14.8703\\
  -61.0000  -14.0200\\
  -60.0000  -13.2111\\
  -59.0000  -12.4466\\
  -58.0000  -11.7293\\
  -57.0000  -11.0624\\
  -56.0000  -10.4494\\
  -55.0000   -9.8940\\
  -54.0000   -9.4009\\
  -53.0000   -8.9748\\
  -52.0000   -8.6222\\
  -51.0000   -8.3503\\
  -50.0000   -8.1683\\
  -49.0000   -8.0879\\
  -48.0000   -8.1243\\
  -47.0000   -8.2980\\
  -46.0000   -8.6377\\
  -45.0000   -9.1851\\
  -44.0000  -10.0046\\
  -43.0000  -11.2035\\
  -42.0000  -12.9821\\
  -41.0000  -15.7850\\
  -40.0000  -20.9985\\
  -39.0000  -44.9784\\
  -38.0000  -20.5399\\
  -37.0000  -13.8070\\
  -36.0000   -9.7195\\
  -35.0000   -6.7141\\
  -34.0000   -4.3105\\
  -33.0000   -2.2962\\
  -32.0000   -0.5584\\
  -31.0000    0.9707\\
  -30.0000    2.3347\\
  -29.0000    3.5640\\
  -28.0000    4.6804\\
  -27.0000    5.7006\\
  -26.0000    6.6374\\
  -25.0000    7.5014\\
  -24.0000    8.3011\\
  -23.0000    9.0439\\
  -22.0000    9.7360\\
  -21.0000   10.3828\\
  -20.0000   10.9887\\
  -19.0000   11.5580\\
  -18.0000   12.0938\\
  -17.0000   12.5990\\
  -16.0000   13.0756\\
  -15.0000   13.5254\\
  -14.0000   13.9494\\
  -13.0000   14.3482\\
  -12.0000   14.7219\\
  -11.0000   15.0704\\
  -10.0000   15.3931\\
   -9.0000   15.6893\\
   -8.0000   15.9580\\
   -7.0000   16.1982\\
   -6.0000   16.4089\\
   -5.0000   16.5891\\
   -4.0000   16.7380\\
   -3.0000   16.8546\\
   -2.0000   16.9384\\
   -1.0000   16.9889\\
         0   17.0057\\
         0   17.0057\\
    1.0000   16.9889\\
    2.0000   16.9384\\
    3.0000   16.8547\\
    4.0000   16.7381\\
    5.0000   16.5893\\
    6.0000   16.4090\\
    7.0000   16.1983\\
    8.0000   15.9581\\
    9.0000   15.6894\\
   10.0000   15.3933\\
   11.0000   15.0706\\
   12.0000   14.7221\\
   13.0000   14.3483\\
   14.0000   13.9495\\
   15.0000   13.5255\\
   16.0000   13.0756\\
   17.0000   12.5990\\
   18.0000   12.0938\\
   19.0000   11.5579\\
   20.0000   10.9885\\
   21.0000   10.3825\\
   22.0000    9.7356\\
   23.0000    9.0435\\
   24.0000    8.3006\\
   25.0000    7.5007\\
   26.0000    6.6366\\
   27.0000    5.6996\\
   28.0000    4.6794\\
   29.0000    3.5628\\
   30.0000    2.3333\\
   31.0000    0.9691\\
   32.0000   -0.5602\\
   33.0000   -2.2983\\
   34.0000   -4.3129\\
   35.0000   -6.7171\\
   36.0000   -9.7234\\
   37.0000  -13.8126\\
   38.0000  -20.5507\\
   39.0000  -44.9679\\
   40.0000  -20.9909\\
   41.0000  -15.7815\\
   42.0000  -12.9802\\
   43.0000  -11.2024\\
   44.0000  -10.0039\\
   45.0000   -9.1848\\
   46.0000   -8.6377\\
   47.0000   -8.2982\\
   48.0000   -8.1245\\
   49.0000   -8.0883\\
   50.0000   -8.1688\\
   51.0000   -8.3509\\
   52.0000   -8.6228\\
   53.0000   -8.9755\\
   54.0000   -9.4015\\
   55.0000   -9.8948\\
   56.0000  -10.4500\\
   57.0000  -11.0631\\
   58.0000  -11.7299\\
   59.0000  -12.4471\\
   60.0000  -13.2115\\
   61.0000  -14.0202\\
   62.0000  -14.8703\\
   63.0000  -15.7593\\
   64.0000  -16.6846\\
   65.0000  -17.6439\\
   66.0000  -18.6351\\
   67.0000  -19.6562\\
   68.0000  -20.7057\\
   69.0000  -21.7828\\
   70.0000  -22.8873\\
   71.0000  -24.0199\\
   72.0000  -25.1827\\
   73.0000  -26.3789\\
   74.0000  -27.6138\\
   75.0000  -28.8944\\
   76.0000  -30.2303\\
   77.0000  -31.6334\\
   78.0000  -33.1193\\
   79.0000  -34.7075\\
   80.0000  -36.4228\\
   81.0000  -38.2975\\
   82.0000  -40.3747\\
   83.0000  -42.7134\\
   84.0000  -45.3993\\
   85.0000  -48.5632\\
   86.0000  -52.4228\\
   87.0000  -57.3829\\
   88.0000  -64.3437\\
   89.0000  -76.0643\\
   90.0000  -87.8294\\
   };
\addlegendentry{AIM}

\addplot[mark size=2pt, mark=x, mark options={solid, red}]
  table[row sep=crcr]{%
  -90.0000  -67.2315\\
  -89.0000  -66.9760\\
  -88.0000  -63.2628\\
  -87.0000  -57.5519\\
  -86.0000  -52.7015\\
  -85.0000  -48.7971\\
  -84.0000  -45.5728\\
  -83.0000  -42.8327\\
  -82.0000  -40.4481\\
  -81.0000  -38.3326\\
  -80.0000  -36.4254\\
  -79.0000  -34.6826\\
  -78.0000  -33.0710\\
  -77.0000  -31.5651\\
  -76.0000  -30.1450\\
  -75.0000  -28.7949\\
  -74.0000  -27.5025\\
  -73.0000  -26.2580\\
  -72.0000  -25.0542\\
  -71.0000  -23.8857\\
  -70.0000  -22.7488\\
  -69.0000  -21.6414\\
  -68.0000  -20.5624\\
  -67.0000  -19.5117\\
  -66.0000  -18.4901\\
  -65.0000  -17.4988\\
  -64.0000  -16.5395\\
  -63.0000  -15.6142\\
  -62.0000  -14.7252\\
  -61.0000  -13.8748\\
  -60.0000  -13.0656\\
  -59.0000  -12.3003\\
  -58.0000  -11.5818\\
  -57.0000  -10.9131\\
  -56.0000  -10.2977\\
  -55.0000   -9.7394\\
  -54.0000   -9.2426\\
  -53.0000   -8.8121\\
  -52.0000   -8.4541\\
  -51.0000   -8.1756\\
  -50.0000   -7.9857\\
  -49.0000   -7.8958\\
  -48.0000   -7.9204\\
  -47.0000   -8.0795\\
  -46.0000   -8.4002\\
  -45.0000   -8.9224\\
  -44.0000   -9.7064\\
  -43.0000  -10.8518\\
  -42.0000  -12.5396\\
  -41.0000  -15.1544\\
  -40.0000  -19.7650\\
  -39.0000  -30.1842\\
  -38.0000  -20.9620\\
  -37.0000  -14.0734\\
  -36.0000   -9.8794\\
  -35.0000   -6.8116\\
  -34.0000   -4.3675\\
  -33.0000   -2.3249\\
  -32.0000   -0.5661\\
  -31.0000    0.9791\\
  -30.0000    2.3558\\
  -29.0000    3.5953\\
  -28.0000    4.7200\\
  -27.0000    5.7468\\
  -26.0000    6.6889\\
  -25.0000    7.5570\\
  -24.0000    8.3598\\
  -23.0000    9.1046\\
  -22.0000    9.7978\\
  -21.0000   10.4447\\
  -20.0000   11.0501\\
  -19.0000   11.6179\\
  -18.0000   12.1517\\
  -17.0000   12.6541\\
  -16.0000   13.1276\\
  -15.0000   13.5737\\
  -14.0000   13.9937\\
  -13.0000   14.3882\\
  -12.0000   14.7577\\
  -11.0000   15.1019\\
  -10.0000   15.4204\\
   -9.0000   15.7125\\
   -8.0000   15.9775\\
   -7.0000   16.2143\\
   -6.0000   16.4220\\
   -5.0000   16.5996\\
   -4.0000   16.7462\\
   -3.0000   16.8611\\
   -2.0000   16.9437\\
   -1.0000   16.9935\\
         0   17.0101\\
         0   17.0101\\
    1.0000   16.9935\\
    2.0000   16.9437\\
    3.0000   16.8611\\
    4.0000   16.7462\\
    5.0000   16.5996\\
    6.0000   16.4220\\
    7.0000   16.2143\\
    8.0000   15.9775\\
    9.0000   15.7125\\
   10.0000   15.4204\\
   11.0000   15.1019\\
   12.0000   14.7577\\
   13.0000   14.3882\\
   14.0000   13.9937\\
   15.0000   13.5737\\
   16.0000   13.1276\\
   17.0000   12.6541\\
   18.0000   12.1517\\
   19.0000   11.6179\\
   20.0000   11.0501\\
   21.0000   10.4447\\
   22.0000    9.7978\\
   23.0000    9.1046\\
   24.0000    8.3598\\
   25.0000    7.5570\\
   26.0000    6.6889\\
   27.0000    5.7468\\
   28.0000    4.7200\\
   29.0000    3.5953\\
   30.0000    2.3558\\
   31.0000    0.9791\\
   32.0000   -0.5661\\
   33.0000   -2.3249\\
   34.0000   -4.3675\\
   35.0000   -6.8116\\
   36.0000   -9.8794\\
   37.0000  -14.0734\\
   38.0000  -20.9620\\
   39.0000  -30.1842\\
   40.0000  -19.7650\\
   41.0000  -15.1544\\
   42.0000  -12.5396\\
   43.0000  -10.8518\\
   44.0000   -9.7064\\
   45.0000   -8.9224\\
   46.0000   -8.4002\\
   47.0000   -8.0795\\
   48.0000   -7.9204\\
   49.0000   -7.8958\\
   50.0000   -7.9857\\
   51.0000   -8.1756\\
   52.0000   -8.4541\\
   53.0000   -8.8121\\
   54.0000   -9.2426\\
   55.0000   -9.7394\\
   56.0000  -10.2977\\
   57.0000  -10.9131\\
   58.0000  -11.5818\\
   59.0000  -12.3003\\
   60.0000  -13.0656\\
   61.0000  -13.8748\\
   62.0000  -14.7252\\
   63.0000  -15.6142\\
   64.0000  -16.5395\\
   65.0000  -17.4988\\
   66.0000  -18.4901\\
   67.0000  -19.5117\\
   68.0000  -20.5624\\
   69.0000  -21.6414\\
   70.0000  -22.7488\\
   71.0000  -23.8857\\
   72.0000  -25.0542\\
   73.0000  -26.2580\\
   74.0000  -27.5025\\
   75.0000  -28.7949\\
   76.0000  -30.1450\\
   77.0000  -31.5651\\
   78.0000  -33.0710\\
   79.0000  -34.6826\\
   80.0000  -36.4254\\
   81.0000  -38.3326\\
   82.0000  -40.4481\\
   83.0000  -42.8327\\
   84.0000  -45.5728\\
   85.0000  -48.7971\\
   86.0000  -52.7015\\
   87.0000  -57.5519\\
   88.0000  -63.2628\\
   89.0000  -66.9760\\
   90.0000  -67.2315\\
   };
\addlegendentry{Proposed}

\end{axis}
\end{tikzpicture}
}}\\
\subfloat[\label{fig:reflectarray_cut2} $\phi = 90^{\circ}$] {\resizebox {0.8\columnwidth} {!} {
\begin{tikzpicture}

\begin{axis}[width=3.0in,
height=1.8in,
scale only axis,
separate axis lines,
every outer x axis line/.append style={black},
every x tick label/.append style={font=\color{black}},
every x tick/.append style={black},
xmin=-90,
xmax=90,
xlabel={$\theta$ [degrees]},
every outer y axis line/.append style={black},
every y tick label/.append style={font=\color{black}},
every y tick/.append style={black},
ymin=-40,
ymax=30,
ylabel={Directivity [dBi]},
axis background/.style={fill=white},
legend style={legend cell align=left, align=left, draw=black,at={(0.42,0.03)},anchor=south}
]

\addplot [color=blue, line width=1pt]
  table[row sep=crcr]{%
  -90.0000  -72.5457 \\
  -89.0000  -42.6953\\
  -88.0000  -36.7676\\
  -87.0000  -33.3502\\
  -86.0000  -30.9888\\
  -85.0000  -29.2269\\
  -84.0000  -27.8630\\
  -83.0000  -26.7918\\
  -82.0000  -25.9533\\
  -81.0000  -25.3121\\
  -80.0000  -24.8485\\
  -79.0000  -24.5535\\
  -78.0000  -24.4274\\
  -77.0000  -24.4797\\
  -76.0000  -24.7311\\
  -75.0000  -25.2178\\
  -74.0000  -26.0016\\
  -73.0000  -27.1908\\
  -72.0000  -28.9944\\
  -71.0000  -31.8899\\
  -70.0000  -37.4328\\
  -69.0000  -56.0365\\
  -68.0000  -35.4476\\
  -67.0000  -29.0971\\
  -66.0000  -25.1307\\
  -65.0000  -22.1849\\
  -64.0000  -19.8191\\
  -63.0000  -17.8348\\
  -62.0000  -16.1257\\
  -61.0000  -14.6288\\
  -60.0000  -13.3037\\
  -59.0000  -12.1234\\
  -58.0000  -11.0696\\
  -57.0000  -10.1292\\
  -56.0000   -9.2932\\
  -55.0000   -8.5559\\
  -54.0000   -7.9135\\
  -53.0000   -7.3642\\
  -52.0000   -6.9083\\
  -51.0000   -6.5477\\
  -50.0000   -6.2859\\
  -49.0000   -6.1285\\
  -48.0000   -6.0834\\
  -47.0000   -6.1611\\
  -46.0000   -6.3752\\
  -45.0000   -6.7429\\
  -44.0000   -7.2845\\
  -43.0000   -8.0198\\
  -42.0000   -8.9552\\
  -41.0000  -10.0406\\
  -40.0000  -11.0573\\
  -39.0000  -11.4726\\
  -38.0000  -10.7231\\
  -37.0000   -8.9804\\
  -36.0000   -6.8830\\
  -35.0000   -4.8314\\
  -34.0000   -2.9571\\
  -33.0000   -1.2777\\
  -32.0000    0.2230\\
  -31.0000    1.5679\\
  -30.0000    2.7787\\
  -29.0000    3.8740\\
  -28.0000    4.8694\\
  -27.0000    5.7783\\
  -26.0000    6.6121\\
  -25.0000    7.3809\\
  -24.0000    8.0935\\
  -23.0000    8.7579\\
  -22.0000    9.3814\\
  -21.0000    9.9703\\
  -20.0000   10.5302\\
  -19.0000   11.0658\\
  -18.0000   11.5809\\
  -17.0000   12.0783\\
  -16.0000   12.5595\\
  -15.0000   13.0255\\
  -14.0000   13.4757\\
  -13.0000   13.9091\\
  -12.0000   14.3239\\
  -11.0000   14.7179\\
  -10.0000   15.0886\\
   -9.0000   15.4334\\
   -8.0000   15.7495\\
   -7.0000   16.0346\\
   -6.0000   16.2863\\
   -5.0000   16.5027\\
   -4.0000   16.6821\\
   -3.0000   16.8230\\
   -2.0000   16.9243\\
   -1.0000   16.9854\\
         0   17.0057\\
         0   17.0057\\
    1.0000   16.9852\\
    2.0000   16.9238\\
    3.0000   16.8222\\
    4.0000   16.6811\\
    5.0000   16.5015\\
    6.0000   16.2850\\
    7.0000   16.0331\\
    8.0000   15.7479\\
    9.0000   15.4316\\
   10.0000   15.0869\\
   11.0000   14.7161\\
   12.0000   14.3222\\
   13.0000   13.9075\\
   14.0000   13.4743\\
   15.0000   13.0244\\
   16.0000   12.5588\\
   17.0000   12.0780\\
   18.0000   11.5811\\
   19.0000   11.0665\\
   20.0000   10.5314\\
   21.0000    9.9721\\
   22.0000    9.3839\\
   23.0000    8.7611\\
   24.0000    8.0972\\
   25.0000    7.3852\\
   26.0000    6.6170\\
   27.0000    5.7837\\
   28.0000    4.8753\\
   29.0000    3.8804\\
   30.0000    2.7855\\
   31.0000    1.5750\\
   32.0000    0.2302\\
   33.0000   -1.2707\\
   34.0000   -2.9507\\
   35.0000   -4.8263\\
   36.0000   -6.8806\\
   37.0000   -8.9831\\
   38.0000  -10.7329\\
   39.0000  -11.4869\\
   40.0000  -11.0696\\
   41.0000  -10.0478\\
   42.0000   -8.9578\\
   43.0000   -8.0190\\
   44.0000   -7.2812\\
   45.0000   -6.7378\\
   46.0000   -6.3686\\
   47.0000   -6.1532\\
   48.0000   -6.0745\\
   49.0000   -6.1186\\
   50.0000   -6.2751\\
   51.0000   -6.5361\\
   52.0000   -6.8958\\
   53.0000   -7.3508\\
   54.0000   -7.8991\\
   55.0000   -8.5406\\
   56.0000   -9.2768\\
   57.0000  -10.1115\\
   58.0000  -11.0505\\
   59.0000  -12.1027\\
   60.0000  -13.2810\\
   61.0000  -14.6037\\
   62.0000  -16.0976\\
   63.0000  -17.8026\\
   64.0000  -19.7812\\
   65.0000  -22.1383\\
   66.0000  -25.0694\\
   67.0000  -29.0054\\
   68.0000  -35.2568\\
   69.0000  -52.8141\\
   70.0000  -37.5414\\
   71.0000  -31.9527\\
   72.0000  -29.0365\\
   73.0000  -27.2218\\
   74.0000  -26.0256\\
   75.0000  -25.2370\\
   76.0000  -24.7468\\
   77.0000  -24.4928\\
   78.0000  -24.4385\\
   79.0000  -24.5629\\
   80.0000  -24.8566\\
   81.0000  -25.3190\\
   82.0000  -25.9591\\
   83.0000  -26.7968\\
   84.0000  -27.8672\\
   85.0000  -29.2303\\
   86.0000  -30.9913\\
   87.0000  -33.3517\\
   88.0000  -36.7674\\
   89.0000  -42.6915\\
   90.0000  -74.7106\\
   };
\addlegendentry{AIM}

\addplot[mark size=2pt, mark=x, mark options={solid, red}]
  table[row sep=crcr]{%
    -90.0000  -55.0976\\
  -89.0000  -42.8320\\
  -88.0000  -36.7915\\
  -87.0000  -33.3101\\
  -86.0000  -30.9124\\
  -85.0000  -29.1279\\
  -84.0000  -27.7490\\
  -83.0000  -26.6676\\
  -82.0000  -25.8223\\
  -81.0000  -25.1769\\
  -80.0000  -24.7112\\
  -79.0000  -24.4160\\
  -78.0000  -24.2918\\
  -77.0000  -24.3483\\
  -76.0000  -24.6066\\
  -75.0000  -25.1040\\
  -74.0000  -25.9034\\
  -73.0000  -27.1154\\
  -72.0000  -28.9514\\
  -71.0000  -31.8812\\
  -70.0000  -37.2334\\
  -69.0000  -43.2078\\
  -68.0000  -33.8637\\
  -67.0000  -28.2481\\
  -66.0000  -24.4931\\
  -65.0000  -21.6417\\
  -64.0000  -19.3272\\
  -63.0000  -17.3737\\
  -62.0000  -15.6842\\
  -61.0000  -14.1997\\
  -60.0000  -12.8824\\
  -59.0000  -11.7066\\
  -58.0000  -10.6547\\
  -57.0000   -9.7142\\
  -56.0000   -8.8764\\
  -55.0000   -8.1357\\
  -54.0000   -7.4885\\
  -53.0000   -6.9332\\
  -52.0000   -6.4699\\
  -51.0000   -6.1006\\
  -50.0000   -5.8288\\
  -49.0000   -5.6604\\
  -48.0000   -5.6031\\
  -47.0000   -5.6680\\
  -46.0000   -5.8699\\
  -45.0000   -6.2283\\
  -44.0000   -6.7686\\
  -43.0000   -7.5224\\
  -42.0000   -8.5236\\
  -41.0000   -9.7849\\
  -40.0000  -11.2046\\
  -39.0000  -12.3047\\
  -38.0000  -12.1120\\
  -37.0000  -10.3657\\
  -36.0000   -7.9897\\
  -35.0000   -5.6651\\
  -34.0000   -3.5832\\
  -33.0000   -1.7525\\
  -32.0000   -0.1405\\
  -31.0000    1.2881\\
  -30.0000    2.5631\\
  -29.0000    3.7088\\
  -28.0000    4.7444\\
  -27.0000    5.6858\\
  -26.0000    6.5459\\
  -25.0000    7.3361\\
  -24.0000    8.0661\\
  -23.0000    8.7445\\
  -22.0000    9.3790\\
  -21.0000    9.9764\\
  -20.0000   10.5427\\
  -19.0000   11.0827\\
  -18.0000   11.6006\\
  -17.0000   12.0994\\
  -16.0000   12.5810\\
  -15.0000   13.0464\\
  -14.0000   13.4955\\
  -13.0000   13.9274\\
  -12.0000   14.3405\\
  -11.0000   14.7327\\
  -10.0000   15.1016\\
   -9.0000   15.4447\\
   -8.0000   15.7594\\
   -7.0000   16.0431\\
   -6.0000   16.2937\\
   -5.0000   16.5092\\
   -4.0000   16.6878\\
   -3.0000   16.8281\\
   -2.0000   16.9290\\
   -1.0000   16.9899\\
         0   17.0101\\
         0   17.0101\\
    1.0000   16.9895\\
    2.0000   16.9284\\
    3.0000   16.8271\\
    4.0000   16.6864\\
    5.0000   16.5075\\
    6.0000   16.2916\\
    7.0000   16.0407\\
    8.0000   15.7565\\
    9.0000   15.4416\\
   10.0000   15.0981\\
   11.0000   14.7288\\
   12.0000   14.3362\\
   13.0000   13.9227\\
   14.0000   13.4905\\
   15.0000   13.0411\\
   16.0000   12.5754\\
   17.0000   12.0935\\
   18.0000   11.5945\\
   19.0000   11.0765\\
   20.0000   10.5364\\
   21.0000    9.9701\\
   22.0000    9.3726\\
   23.0000    8.7381\\
   24.0000    8.0596\\
   25.0000    7.3295\\
   26.0000    6.5391\\
   27.0000    5.6787\\
   28.0000    4.7370\\
   29.0000    3.7008\\
   30.0000    2.5543\\
   31.0000    1.2780\\
   32.0000   -0.1525\\
   33.0000   -1.7674\\
   34.0000   -3.6028\\
   35.0000   -5.6924\\
   36.0000   -8.0293\\
   37.0000  -10.4233\\
   38.0000  -12.1828\\
   39.0000  -12.3631\\
   40.0000  -11.2384\\
   41.0000   -9.8013\\
   42.0000   -8.5305\\
   43.0000   -7.5237\\
   44.0000   -6.7665\\
   45.0000   -6.2240\\
   46.0000   -5.8641\\
   47.0000   -5.6609\\
   48.0000   -5.5948\\
   49.0000   -5.6511\\
   50.0000   -5.8188\\
   51.0000   -6.0896\\
   52.0000   -6.4580\\
   53.0000   -6.9203\\
   54.0000   -7.4746\\
   55.0000   -8.1207\\
   56.0000   -8.8602\\
   57.0000   -9.6966\\
   58.0000  -10.6355\\
   59.0000  -11.6857\\
   60.0000  -12.8593\\
   61.0000  -14.1740\\
   62.0000  -15.6551\\
   63.0000  -17.3404\\
   64.0000  -19.2879\\
   65.0000  -21.5933\\
   66.0000  -24.4292\\
   67.0000  -28.1516\\
   68.0000  -33.6645\\
   69.0000  -42.4499\\
   70.0000  -37.1744\\
   71.0000  -31.8872\\
   72.0000  -28.9579\\
   73.0000  -27.1178\\
   74.0000  -25.9014\\
   75.0000  -25.0980\\
   76.0000  -24.5969\\
   77.0000  -24.3350\\
   78.0000  -24.2748\\
   79.0000  -24.3952\\
   80.0000  -24.6862\\
   81.0000  -25.1472\\
   82.0000  -25.7870\\
   83.0000  -26.6257\\
   84.0000  -27.6985\\
   85.0000  -29.0657\\
   86.0000  -30.8335\\
   87.0000  -33.2042\\
   88.0000  -36.6345\\
   89.0000  -42.5427\\
   90.0000  -56.1840\\
   };
\addlegendentry{Proposed}

\end{axis}
\end{tikzpicture}
}}\\
\subfloat[\label{fig:reflectarray_cut3} $\phi = 45^{\circ}$] {\resizebox {0.8\columnwidth} {!} {\begin{tikzpicture}

\begin{axis}[width=3.0in,
height=1.8in,
scale only axis,
separate axis lines,
every outer x axis line/.append style={black},
every x tick label/.append style={font=\color{black}},
every x tick/.append style={black},
xmin=-90,
xmax=90,
xlabel={$\theta$ [degrees]},
every outer y axis line/.append style={black},
every y tick label/.append style={font=\color{black}},
every y tick/.append style={black},
ymin=-40,
ymax=30,
ylabel={Directivity [dBi]},
axis background/.style={fill=white},
legend style={legend cell align=left, align=left, draw=black,at={(0.42,0.03)},anchor=south}
]

\addplot [color=blue, line width=1pt]
  table[row sep=crcr]{%
  -90.0000  -76.3477 \\
  -89.0000  -55.9771 \\
  -88.0000  -49.8313\\
  -87.0000  -46.4469\\
  -86.0000  -44.2253\\
  -85.0000  -42.6833\\
  -84.0000  -41.6107\\
  -83.0000  -40.8949\\
  -82.0000  -40.4592\\
  -81.0000  -40.2256\\
  -80.0000  -40.0772\\
  -79.0000  -39.8194\\
  -78.0000  -39.1923\\
  -77.0000  -38.0109\\
  -76.0000  -36.3284\\
  -75.0000  -34.3709\\
  -74.0000  -32.3513\\
  -73.0000  -30.3927\\
  -72.0000  -28.5485\\
  -71.0000  -26.8344\\
  -70.0000  -25.2500\\
  -69.0000  -23.7887\\
  -68.0000  -22.4421\\
  -67.0000  -21.2024\\
  -66.0000  -20.0626\\
  -65.0000  -19.0169\\
  -64.0000  -18.0606\\
  -63.0000  -17.1900\\
  -62.0000  -16.4025\\
  -61.0000  -15.6960\\
  -60.0000  -15.0690\\
  -59.0000  -14.5204\\
  -58.0000  -14.0491\\
  -57.0000  -13.6537\\
  -56.0000  -13.3322\\
  -55.0000  -13.0805\\
  -54.0000  -12.8922\\
  -53.0000  -12.7564\\
  -52.0000  -12.6556\\
  -51.0000  -12.5641\\
  -50.0000  -12.4464\\
  -49.0000  -12.2590\\
  -48.0000  -11.9578\\
  -47.0000  -11.5095\\
  -46.0000  -10.9040\\
  -45.0000  -10.1584\\
  -44.0000   -9.3095\\
  -43.0000   -8.4014\\
  -42.0000   -7.4740\\
  -41.0000   -6.5574\\
  -40.0000   -5.6710\\
  -39.0000   -4.8243\\
  -38.0000   -4.0193\\
  -37.0000   -3.2518\\
  -36.0000   -2.5127\\
  -35.0000   -1.7896\\
  -34.0000   -1.0680\\
  -33.0000   -0.3337\\
  -32.0000    0.4255\\
  -31.0000    1.2175\\
  -30.0000    2.0447\\
  -29.0000    2.9035\\
  -28.0000    3.7855\\
  -27.0000    4.6793\\
  -26.0000    5.5722\\
  -25.0000    6.4524\\
  -24.0000    7.3095\\
  -23.0000    8.1353\\
  -22.0000    8.9239\\
  -21.0000    9.6715\\
  -20.0000   10.3760\\
  -19.0000   11.0370\\
  -18.0000   11.6549\\
  -17.0000   12.2312\\
  -16.0000   12.7678\\
  -15.0000   13.2669\\
  -14.0000   13.7310\\
  -13.0000   14.1623\\
  -12.0000   14.5625\\
  -11.0000   14.9331\\
  -10.0000   15.2752\\
   -9.0000   15.5890\\
   -8.0000   15.8744\\
   -7.0000   16.1306\\
   -6.0000   16.3566\\
   -5.0000   16.5511\\
   -4.0000   16.7126\\
   -3.0000   16.8399\\
   -2.0000   16.9318\\
   -1.0000   16.9872\\
         0   17.0057\\
         0   17.0057\\
    1.0000   16.9871\\
    2.0000   16.9314\\
    3.0000   16.8394\\
    4.0000   16.7120\\
    5.0000   16.5503\\
    6.0000   16.3557\\
    7.0000   16.1295\\
    8.0000   15.8732\\
    9.0000   15.5877\\
   10.0000   15.2737\\
   11.0000   14.9315\\
   12.0000   14.5606\\
   13.0000   14.1602\\
   14.0000   13.7288\\
   15.0000   13.2646\\
   16.0000   12.7652\\
   17.0000   12.2284\\
   18.0000   11.6519\\
   19.0000   11.0337\\
   20.0000   10.3725\\
   21.0000    9.6676\\
   22.0000    8.9196\\
   23.0000    8.1307\\
   24.0000    7.3045\\\\
   25.0000    6.4470\\
   26.0000    5.5664\\
   27.0000    4.6730\\
   28.0000    3.7789\\
   29.0000    2.8967\\
   30.0000    2.0379\\
   31.0000    1.2109\\
   32.0000    0.4193\\
   33.0000   -0.3393\\
   34.0000   -1.0727\\
   35.0000   -1.7933\\
   36.0000   -2.5153\\\\
   37.0000   -3.2532\\
   38.0000   -4.0194\\
   39.0000   -4.8228\\
   40.0000   -5.6678\\
   41.0000   -6.5523\\
   42.0000   -7.4667\\
   43.0000   -8.3917\\
   44.0000   -9.2971\\
   45.0000  -10.1431\\
   46.0000  -10.8861\\
   47.0000  -11.4895\\
   48.0000  -11.9366\\
   49.0000  -12.2376\\
   50.0000  -12.4255\\
   51.0000  -12.5445\\
   52.0000  -12.6374\\
   53.0000  -12.7397\\
   54.0000  -12.8771\\
   55.0000  -13.0668\\
   56.0000  -13.3198\\
   57.0000  -13.6426\\
   58.0000  -14.0391\\
   59.0000  -14.5115\\
   60.0000  -15.0612\\
   61.0000  -15.6893\\
   62.0000  -16.3969\\
   63.0000  -17.1856\\
   64.0000  -18.0573\\
   65.0000  -19.0150\\
   66.0000  -20.0622\\
   67.0000  -21.2037\\
   68.0000  -22.4454\\
   69.0000  -23.7942\\
   70.0000  -25.2581\\
   71.0000  -26.8455\\
   72.0000  -28.5630\\
   73.0000  -30.4108\\
   74.0000  -32.3724\\
   75.0000  -34.3926\\
   76.0000  -36.3447\\
   77.0000  -38.0120\\
   78.0000  -39.1686\\
   79.0000  -39.7695\\
   80.0000  -40.0074\\
   81.0000  -40.1430\\
   82.0000  -40.3677\\
   83.0000  -40.7954\\
   84.0000  -41.5017\\
   85.0000  -42.5610\\
   86.0000  -44.0821\\
   87.0000  -46.2673\\
   88.0000  -49.5768\\
   89.0000  -55.4941\\
   90.0000  -83.7756\\
   };
\addlegendentry{AIM}

\addplot[mark size=2pt, mark=x, mark options={solid, red}]
  table[row sep=crcr]{%
  -90.0000  -65.2365\\
  -89.0000  -57.7582\\
  -88.0000  -50.3498\\
  -87.0000  -46.5765\\
  -86.0000  -44.1496\\
  -85.0000  -42.4622\\
  -84.0000  -41.2619\\
  -83.0000  -40.4152\\
  -82.0000  -39.8316\\
  -81.0000  -39.4262\\
  -80.0000  -39.0900\\
  -79.0000  -38.6694\\
  -78.0000  -37.9806\\
  -77.0000  -36.8910\\
  -76.0000  -35.4103\\
  -75.0000  -33.6731\\
  -74.0000  -31.8365\\
  -73.0000  -30.0131\\
  -72.0000  -28.2639\\
  -71.0000  -26.6158\\
  -70.0000  -25.0770\\
  -69.0000  -23.6471\\
  -68.0000  -22.3219\\
  -67.0000  -21.0961\\
  -66.0000  -19.9645\\
  -65.0000  -18.9223\\
  -64.0000  -17.9653\\
  -63.0000  -17.0904\\
  -62.0000  -16.2946\\
  -61.0000  -15.5758\\
  -60.0000  -14.9319\\
  -59.0000  -14.3612\\
  -58.0000  -13.8617\\
  -57.0000  -13.4308\\
  -56.0000  -13.0649\\
  -55.0000  -12.7586\\
  -54.0000  -12.5040\\
  -53.0000  -12.2889\\
  -52.0000  -12.0967\\
  -51.0000  -11.9045\\
  -50.0000  -11.6842\\
  -49.0000  -11.4050\\
  -48.0000  -11.0384\\
  -47.0000  -10.5652\\
  -46.0000   -9.9818\\
  -45.0000   -9.3008\\
  -44.0000   -8.5472\\
  -43.0000   -7.7510\\
  -42.0000   -6.9408\\
  -41.0000   -6.1393\\
  -40.0000   -5.3622\\
  -39.0000   -4.6177\\
  -38.0000   -3.9075\\
  -37.0000   -3.2270\\
  -36.0000   -2.5669\\
  -35.0000   -1.9135\\
  -34.0000   -1.2506\\
  -33.0000   -0.5618\\
  -32.0000    0.1667\\
  -31.0000    0.9438\\
  -30.0000    1.7708\\
  -29.0000    2.6420\\
  -28.0000    3.5457\\
  -27.0000    4.4668\\
  -26.0000    5.3897\\
  -25.0000    6.2998\\
  -24.0000    7.1853\\
  -23.0000    8.0370\\
  -22.0000    8.8485\\
  -21.0000    9.6159\\
  -20.0000   10.3372\\
  -19.0000   11.0121\\
  -18.0000   11.6415\\
  -17.0000   12.2268\\
  -16.0000   12.7704\\
  -15.0000   13.2748\\
  -14.0000   13.7425\\
  -13.0000   14.1760\\
  -12.0000   14.5774\\
  -11.0000   14.9483\\
  -10.0000   15.2900\\
   -9.0000   15.6029\\
   -8.0000   15.8869\\
   -7.0000   16.1417\\
   -6.0000   16.3662\\
   -5.0000   16.5592\\
   -4.0000   16.7195\\
   -3.0000   16.8457\\
   -2.0000   16.9368\\
   -1.0000   16.9918\\
         0   17.0101\\
         0   17.0101\\
    1.0000   16.9916\\
    2.0000   16.9364\\
    3.0000   16.8451\\
    4.0000   16.7187\\
    5.0000   16.5582\\
    6.0000   16.3650\\
    7.0000   16.1402\\
    8.0000   15.8853\\
    9.0000   15.6010\\
   10.0000   15.2879\\
   11.0000   14.9461\\
   12.0000   14.5749\\
   13.0000   14.1733\\
   14.0000   13.7395\\
   15.0000   13.2715\\
   16.0000   12.7669\\
   17.0000   12.2230\\
   18.0000   11.6373\\
   19.0000   11.0077\\
   20.0000   10.3323\\
   21.0000    9.6105\\
   22.0000    8.8425\\
   23.0000    8.0304\\
   24.0000    7.1780\\
   25.0000    6.2917\\
   26.0000    5.3806\\
   27.0000    4.4568\\
   28.0000    3.5348\\
   29.0000    2.6303\\
   30.0000    1.7584\\
   31.0000    0.9310\\
   32.0000    0.1539\\
   33.0000   -0.5743\\
   34.0000   -1.2625\\
   35.0000   -1.9244\\
   36.0000   -2.5768\\
   37.0000   -3.2357\\
   38.0000   -3.9148\\
   39.0000   -4.6236\\
   40.0000   -5.3666\\
   41.0000   -6.1421\\
   42.0000   -6.9417\\
   43.0000   -7.7499\\
   44.0000   -8.5436\\
   45.0000   -9.2945\\
   46.0000   -9.9724\\
   47.0000  -10.5528\\
   48.0000  -11.0232\\
   49.0000  -11.3874\\
   50.0000  -11.6648\\
   51.0000  -11.8837\\
   52.0000  -12.0751\\
   53.0000  -12.2670\\
   54.0000  -12.4818\\
   55.0000  -12.7366\\
   56.0000  -13.0430\\
   57.0000  -13.4092\\
   58.0000  -13.8405\\
   59.0000  -14.3405\\
   60.0000  -14.9118\\
   61.0000  -15.5564\\
   62.0000  -16.2762\\
   63.0000  -17.0732\\
   64.0000  -17.9497\\
   65.0000  -18.9087\\
   66.0000  -19.9535\\
   67.0000  -21.0886\\
   68.0000  -22.3189\\
   69.0000  -23.6500\\
   70.0000  -25.0877\\
   71.0000  -26.6368\\
   72.0000  -28.2985\\
   73.0000  -30.0649\\
   74.0000  -31.9086\\
   75.0000  -33.7657\\
   76.0000  -35.5154\\
   77.0000  -36.9871\\
   78.0000  -38.0369\\
   79.0000  -38.6619\\
   80.0000  -39.0115\\
   81.0000  -39.2804\\
   82.0000  -39.6231\\
   83.0000  -40.1436\\
   84.0000  -40.9193\\
   85.0000  -42.0289\\
   86.0000  -43.5859\\
   87.0000  -45.7955\\
   88.0000  -49.1191\\
   89.0000  -55.0160\\
   90.0000  -71.0839\\
   };
\addlegendentry{Proposed}

\end{axis}
\end{tikzpicture}}}
\end{center}
\caption{Directivity of the $11 \times 11$ reflectarray considered in Sec.~\ref{sec:meanderreflectarray} calculated with AIM and the proposed technique.}
\label{fig:directivity_meander}
\end{figure}

\begin{table}[t]
\caption{Simulation Settings and Results for the $11 \times 11$ reflectarray considered in Sec.~\ref{sec:meanderreflectarray}}
\label{tab:reflectarray_meander}
\begin{center}
\begin{tabular}{|l| c | c|}
\hline
& AIM & Proposed \\
\hline
\multicolumn{3}{|c|}{AIM Parameters}\\ \hline
Number of stencils in $x$ dir. & $126 $ & $84$ \\
Number of stencils in $y$ dir. & $126 $ & $84$ \\
Number of stencils in $z$ dir. & $2$ & $2$\\
Interpolation order & 3 & 3 \\ 
Number of near-field stencils & 4 &4  \\
\hline
\multicolumn{3}{|c|}{Memory Consumption}\\
\hline
Total number of unknowns & 262,616 & 27,588 \\
Memory used & 67~GB & 5.4~GB \\
\hline
\multicolumn{3}{|c|}{Timing Results}\\
\hline
Macromodel generation & N/A & 14.2~min \\
Matrix fill time & 5.31~h &  7.8~min \\
Preconditioner factorization & 3.83~h & 4.15~min \\
Iterative solver & 2.04~h & 26~s \\
Total computation time & 11.18~h & 27~min \\
\hline
\end{tabular}
\end{center}
\end{table}

As a final example, we consider a single-layer reflectarray with very fine meander-line features~\cite{Qin2015}.
The top view of this reflectarray is shown in Fig.~\ref{fig:reflectarray_meander}.
The mesh size of such elements is electrically very small, which causes conditioning issues with the standard MoM. 
However, with the proposed approach the structure can be simulated faster due to fewer unknowns and better conditioning of the equations to be solved.

All elements in this example are suspended in free space $0.75~\mathrm{mm}$ above a PEC ground plane.
As in Sec.~\ref{sec:jerusalemcross}, we use the image theory to model the ground plane.
The reflectarray has a total of $11 \times 11$ elements, out of which eight elements are unique.
The structure is designed to scatter fields with the main beam in the broadside direction.
The reflectarray is excited by a dipole antenna operating at $f = 14~\mathrm{GHz}$ that is placed $43~{\rm mm}$ along the axis of the reflectarray so that focal length to diameter ratio is 0.65.
We simulated the problem with the in-house MoM code and the proposed macromodeling technique, both accelerated with AIM.
In the proposed method, an equivalent box of dimensions $5.5~\mathrm{mm} \times 5.5~\mathrm{mm} \times 3.0~\mathrm{mm}$ was introduced to enclose each element.

Fig.~\ref{fig:directivity_meander} shows the directivity of the reflectarray calculated with both techniques.
The agreement between the results obtained with the proposed method and the standard MoM code confirms that proposed method can accurately capture the strong coupling between the elements and fine features of the reflectarray unit cell.

Tab.~\ref{tab:reflectarray_meander} shows AIM parameters, storage statistics, and timing statistics to simulate this problem.
As seen from Tab.~\ref{tab:reflectarray_meander}, the proposed method is $24$ times faster and consumes $12$ times less memory than the standard MoM solver.
The proposed method solves the problem in $1/2~{\rm h}$ as opposed to $11~{\rm h}$ required with the standard MoM formulation, which is a significant savings.
The proposed method is faster because it requires solving a problem with 9 times less number of unknowns. 
Furthermore, the proposed formulation converges significantly faster than the standard MoM formulation due to a significantly smaller condition number.
We computed, in PETSc, the condition number of the proposed formulation and the standard MoM formulation for a smaller sized reflectarray with $3 \times 3$ elements\footnote{Computing the condition number of the $11 \times 11$ reflectarray was not feasible due to its prohibitive computational cost.}. It was found that the condition number of the proposed formulation was 129.31, which was significantly smaller than the condition number of the standard MoM equations which was $1.013 \times 10^5$.

This example demonstrates that the proposed method can be very efficient, in terms of computation time and memory consumption, to simulate arrays with complex elements.

\section{Conclusions}
\label{sec:conclusion}
In the classical circuit theory, complexity of a large electrical network of linear elements is often
reduced using the Norton equivalent circuit models. 
With the Norton equivalent circuit model, a complex portion of a large electrical network can be replaced by an equivalent current source and impedance. 
By doing this, the equivalent electrical network can be greatly simplified, as it will have fewer nodes, branches, and elements. 
In this paper, we explored whether this idea can be extended to modeling electromagnetic scatterers. 
As demonstrated in this work, this is indeed possible by combining the equivalence theorem and Stratton-Chu formulation. 
It was also found that both RWG and Dual RWG basis functions were necessary for robust numerical implementation. We presented an equivalent model, which we referred to as a macromodel, through which a complex scatterer is
replaced by an equivalent electric current source, which is analogous to the Norton equivalent current
source, and a transfer operator, which is analogous to the Norton impedance. 
Like the Norton theorem, the macromodel approach presented in this paper makes no heuristic approximations and is exact.

In this paper, we applied this macromodeling technique to efficiently simulate large arrays of complex
scatterers. 
The proposed method is more efficient than the standard MoM for three reasons. 
First, it significantly reduces unknowns count, which results in lower solution time and matrix fill time. 
Second, the proposed method results in a linear system with better conditioning than the standard MoM, leading to faster convergence of iterative
methods when applied to multiscale problems. 
Third, the proposed method exploits the repeatability of
elements in large arrays. The proposed method was applied to compute scattering from an array of
spherical helix antennas and reflectarrays.
It was shown that the method is up to 20 times faster and
consumes up to 12 times less memory than the standard MoM simulation, while giving accurate results.

Although in this work we investigated array problems only, the proposed macromodel approach can be
applied to many other multiscale problems. 
For example, it can be applied to model antennas for channel modeling or model antennas on large aircraft or ships.
The proposed approach can also be adopted to model uncertainty, when many runs are needed to capture statistics of a problem.
Currently, we are working to extend the proposed method to include dielectric substrates.
Through this extension, we will be able to simulate many other practical metasurfaces and reflectarrays.

\section{Acknowledgment}
The authors thank  IDS Corporation for providing a license for the Antenna Design Framework (ADF).  
The authors also thank Mr. Ciaran Geaney for his help with the reflectarray test cases.
\bibliographystyle{IEEEtran}
\bibliography{IEEEabrv,biblio}

\end{document}